%% file: DG750.tex
\documentclass[12pt]{article}
\pdfoutput=1

\usepackage{amsmath,amsfonts,amssymb,epsfig}
\usepackage{slashed}
\usepackage{cite}
\usepackage[width=0.9\textwidth,margin={30pt,30pt},font=small,labelfont=bf]{caption}
\numberwithin{equation}{section}

\usepackage{hyperref}

\usepackage{xcolor}
\hypersetup{
    colorlinks,
    linkcolor={red!0!black},
    citecolor={blue!50!black},
    urlcolor={blue!80!black}
}

\usepackage{graphicx}
\usepackage{cancel}
\usepackage{cite}
\usepackage{color}
\usepackage{bm}
\usepackage{multirow}

\usepackage{xspace}

\def\SARAH{{\tt SARAH}\xspace}
\newcommand{\SPheno}{{\tt SPheno}\xspace}

\def\hE{{\hat{E}}}
\def\hEp{{\hat{E}^{\prime}}}
\def\hfE{{\hat{\tilde{E}}}}
\def\hfEp{{\hat{\tilde{E}}^{\prime}}}

\def\s2b{s_{2\beta}}
\def\c2b{c_{2\beta}}

\def\G{\Gamma}
\def\g{\gamma}
\def\gp{{g_Y}{}}

\def\ov{\overline}
\def\tr{\mathrm{tr}}

\def\ms{M_S}

\def\fb{\ \mathrm{fb}}

\def\TeV{\ \mathrm{TeV}}

\def\Stogg{S \rightarrow \gamma \gamma}
\def\Stogluglu{S \rightarrow g g }

\def\beq{\begin{equation}}
\def\eeq{\end{equation}}
\def\bal{\begin{align}}
\def\eal{\end{align}}

\newcommand{\scr}[1]{\ensuremath{\mathcal{#1}}}
\newcommand{\exclude}[1]{}
\def\nn{\nonumber}

\def\Ra{$\mathbf{R_a} \ $}
\def\Rb{$\mathbf{R_b} \ $}

\def\Rva{$\bm {\slashed R_a} \ $}
\def\Rvb{$\bm {\slashed R_b} \ $}

\def\nn{\nonumber}

\begin{document}

\begin{flushright}
\end{flushright}
\begin{center}

\vspace{1cm}
{\LARGE{\bf The Di-Photon Excess in a Perturbative \\[0.5em]
SUSY Model
 }}

\vspace{1cm}

\large{  Karim Benakli$^\clubsuit$ \let\thefootnote\relax\footnote{$^\clubsuit$kbenakli@lpthe.jussieu.fr},
Luc Darm\' e$^\heartsuit$ \footnote{$^\heartsuit$darme@lpthe.jussieu.fr},
Mark~D.~Goodsell$^\diamondsuit$ \footnote{$^\diamondsuit$goodsell@lpthe.jussieu.fr} and
Julia  Harz$^\spadesuit$ \footnote{$^\spadesuit$jharz@lpthe.jussieu.fr}
 \\[5mm]}

{\small
\emph{1-- Sorbonne Universit\'es, UPMC Univ Paris 06, UMR 7589, LPTHE, F-75005, Paris, France \\
2-- CNRS, UMR 7589, LPTHE, F-75005, Paris, France \\
}}

\end{center}
\vspace{0.7cm}

\abstract{We show that a 750 GeV di-photon  excess as reported by the ATLAS and CMS experiments can be reproduced by the  Minimal Dirac Gaugino Supersymmetric Standard Model (MDGSSM) without the need of any ad-hoc addition of new states. The scalar resonance is identified with the spin-0 partner of the Dirac bino. We perform a thorough analysis of constraints coming from the mixing of the scalar with the Higgs boson, the stability of the vacuum and the requirement of perturbativity of the couplings up to very high energy scales. We exhibit examples of regions of the parameter space that respect all the constraints while reproducing the excess. We point out how trilinear couplings that are expected to arise in supersymmetry-breaking mediation scenarios, but were ignored in the previous literature on the subject, play an important role. }

\newpage

\tableofcontents

\setcounter{footnote}{0}

\section{Introduction}
\label{SEC:INTRO}

\input{Intro}

% \section{The Di-Photon Excess}
% \label{SEC:EXCESS}
% 
% \input{Subsections/diphoton}
% 
%  
 
\section{The Minimal Dirac Gaugino Model}
\label{SEC:MDGSSM}

\subsection{Model Content and Lagrangian}

\input{ModelContent}

\section{Generating Trilinear and Quartic Couplings}
\label{SEC:TRILINEARS}

\input{Trilinears}

% \subsection{Benchmark scenarios}
% 
% \input{Subsections/benchmark}

\section{Constraining the MDGSSM from the diphoton excess}
\label{SEC:CONSTRAINTS}

We analyse in this section various theoretical and experimental constraints lying on the general model presented above. We start by considering the basics of production and decay of the scalar singlet and then study the most relevant collider constraints on our model. We finally investigate the requirements we need to impose in order to remain perturbative up to the GUT scale and avoid the appearance of Charge or Colour Breaking Vacuua.

\subsection{Production and Decay in the MDGSSM}
\input{productionMDGSSM}

\subsection{Constraints from Higgs mass mixing and 8 TeV data}
\label{sec:mH}

\input{Mixing}

\subsection{Bounds on colour octets}
\input{Octets}

\subsection{Perturbativity and Landau Poles}
\label{sec:Landau}
\input{Unification}

\subsection{Vacuum stability}
\label{sec:vacua}

\input{VacuumStudy}

\section{Finding a Di-Photon excess in the MDGSSM}
\label{SEC:EXCESS}

\subsection{Prelude}
\label{sec:prelude}

\input{Prelude}

\subsection{R-Symmetry conserving Scenarios}

\input{Pheno_Rsym}

\subsection{R-Symmetry violating Scenarios}
\input{Pheno_Rviolating}

\section{Conclusions}
\label{SEC:Conclusions}
\input{Conclusions}

\section*{Acknowledgments}

% MDG and KB acknowledge support from Agence Nationale de Recherche grant ANR-15-CE31-0002 “HiggsAutomator.”
%  Add for ILP and ERC-LHC 
The work of K.B and M.D.G is supported by the Agenge Nationale de Recherche under grant ANR-15-CE31-0002 ``HiggsAutomator''. K.B also acknowledges the support of the ERC grant Higgs@LHC. The work of L.D and J.H is supported by the Labex ``Institut Lagrange de Paris''. 

%\newpage

%\appendix

% \section{Appendices}
% \addcontentsline{toc}{section}{Appendix}

% \begin{Appendix}
% \label{sec:appendix}

% \subsection{Implementation}
% \label{APP:implementation}

%  RGEs are then solved iteratively (using numerical routines from SPheno~\cite{porod_spheno_2003,porod_spheno_2012}) until we reach a solution satisfying both boundary conditions at the required precision. 

% The RGEs above the SUSY scale have been obtained using the public code \SARAH~(see ref.~\cite{staub_sarah_2008,staub_automatic_2011,staub_superpotential_2010,staub_sarah_2013,staub_sarah_2014} and ref.~\cite{goodsell_two-loop_2013}).  

% % 
% % \subsection{Old summary}
% % \label{APP:OldSummary}
% % 
% % \input{Subsections/oldsummary}
% % 

% \end{Appendix}

% 
%\begin{thebibliography}{99}
\newpage

\bibliographystyle{unsrt}
\bibliography{DG750}

\end{document}

%% file: Intro.tex
In the first presentation of LHC Run 2 data, both experiments ATLAS and CMS presented an excess in the di-photon mass spectrum for comparable invariant masses. The CMS analysis 
observed its largest excess in the di-photon mass spectrum based on $2.6~\mathrm{fb^{-1}}$ of pp collisions at $\sqrt{s}= 13~\mathrm{TeV}$ for an invariant mass of $760~\mathrm{GeV}$ with a local significance of $2.6~\sigma$ and a global significance of smaller than $1.2~\sigma$  \cite{CMS-PAS-EXO-15-004}. Similarly, the ATLAS collaboration reported the largest deviation from the background hypothesis for an invariant mass of $750~\mathrm{GeV}$ using $3.2~\mathrm{fb^{-1}}$ of data, leading to a local significance of $3.6~\sigma$ and a global significance of $2.0~\sigma$ taking into account the look-elsewhere-effect in the mass range of $m_{\gamma\gamma} \in [200-2000]~\mathrm{GeV}$ \cite{ATLAS-CONF-2015-081}. 

After updating and refining their analysis, CMS achieved an improved sensitivity by more than $20~\%$ and added a new data set which was taken with $B=0~\mathrm{T}$ reaching as well a comparable $3.3~\mathrm{fb^{-1}}$. The modest excess at $750~\mathrm{GeV}$ for the combined 8 and 13 TeV data remained with $3.4~\sigma$ (local) and $1.6~\sigma$ (global) significance \cite{CMS-PAS-EXO-16-018}. ATLAS updated their $8~\mathrm{TeV}$ analysis and confirmed the modest excess at $750~\mathrm{GeV}$ in the Run I data set with a significance of $1.9~\sigma$. %\julia{conf note not yet out}. 
Thus, the recent updates strengthen the hint for a new physics signal.

For the Spin-0 hypothesis and under the assumption of $\Gamma/\ms = 0.014 \times 10^{-2}$ (with $\ms$ the scalar singlet mass) the combined dataset of CMS with $3.3~\mathrm{fb^{-1}} (13~\mathrm{TeV})$ and $19.7~\mathrm{fb^{-1}} (8~\mathrm{TeV})$ gives the production cross-section times branching ratio into two photons to be 
\begin{align}
\sigma^{13~\mathrm{TeV}} \cdot B_{\gamma \gamma} \approx 3.7 \pm  2 \mathrm{fb}.
\end{align}
while one analysis of the ATLAS data gives \cite{Buckley:2016mbr}.
\begin{align}
\sigma^{13~\mathrm{TeV}} \cdot B_{\gamma \gamma} \approx 12 \pm  2 \mathrm{fb}.
\end{align}

An interpretation of this excess is that it is due to the production and subsequent
decay of a scalar resonance of mass $750$ GeV; while there have been many alternatives proposed (too many to mention here), we shall restrict to that case here as the most obvious and least tuned option in perturbative theories. The existence of such a particle with a mass close to the electroweak scale implies a new hierarchy problem that cannot obviously have an anthropic explanation, and this naturally strengthens the case for low-energy supersymmetry. However, the observed rate of diphoton production via the resonance is too large compared to what is expected from a heavy Higgs companion
of the light Standard Model (SM)-like one, and in particular it is very difficult to justify in the Minimal Supersymmetric Standard Model (MSSM) (see e.g. \cite{Angelescu:2015uiz}\footnote{Note that although there have been several attempts to fit the excess in just the MSSM, such as in \cite{Choudhury:2016jbc,Djouadi:2016oey}, they require a large fine-tuning of masses/parameters to be on resonance, and even then there remain questions about the viability of the scenario from e.g. vacuum stability constraints or sufficient production.}). In fact, the
interpretation of the excess is challenging for most previously proposed
supersymmetric extensions of the Standard Model, and of the perturbative models proposed since the announcement almost all invoke additional vector-like fermions and/or bosons. For an early review see \cite{Staub:2016dxq}. In this work we shall show, on the other hand, that a previously proposed supersymmetric extension of the Standard Model called the Minimal Dirac Gaugino Supersymmetric Standard Model (MDGSSM) \cite{Benakli:2014cia} contains all of the ingredients to explain the excess.

Since the proposal in \cite{fayet_massive_1978}  of extending the MSSM with extra states in the adjoint representation of the Standard Model to allow Dirac gaugino masses, this possibility has been subject to many studies due to their theoretical and phenomenological advantages: they allow simpler models of supersymmetry-breaking due to preserving an R-symmetry; their masses are supersoft \cite{fox_dirac_2002} and supersafe from collider searches \cite{Heikinheimo:2011fk,Kribs:2012gx,Kribs:2013oda}; they ameliorate the SUSY flavour problem \cite{Kribs:2007ac,Fok:2010vk,Dudas:2013gga}; and contain new couplings which aid the naturalness of the Higgs mass \cite{Belanger:2009wf,Benakli:2011kz,Benakli:2012cy,Benakli:2014cia,Bertuzzo:2014bwa}. Indeed, multiple realisations have been proposed that differ by the fate of R-symmetry, the presence or absence of additional states and interactions \cite{Polchinski:1982an,Hall:1990hq,fox_dirac_2002,Nelson:2002ca,Antoniadis:2006uj,%
Amigo:2008rc,Plehn:2008ae,Benakli:2008pg,Benakli:2009mk,Choi:2009ue,Benakli:2010gi,Choi:2010gc,%
Carpenter:2010as,Kribs:2010md,Abel:2011dc,Davies:2011mp,Benakli:2011kz,Kalinowski:2011zzc,Frugiuele:2011mh,%
Bertuzzo:2012su,Davies:2012vu,Argurio:2012cd,Argurio:2012bi,Frugiuele:2012pe,%
Frugiuele:2012kp,Benakli:2012cy,Itoyama:2013sn,Chakraborty:2013gea,Csaki:2013fla,Itoyama:2013vxa,Beauchesne:2014pra,%
Bertuzzo:2014bwa,Goodsell:2014dia,Busbridge:2014sha,Chakraborty:2014sda,Ding:2015wma,Alves:2015kia,Alves:2015bba,Carpenter:2015mna,Martin:2015eca} (for a short introduction see for example \cite{Benakli:2011vb}). Here we consider the case of the MDGSSM which was introduced with a minimal content of extra states to automatically preserve unification of gauge couplings while allowing the new couplings to the Higgs to enhance naturalness and allow the boundary conditions to be unified at a high energy scale.

We will show that it is  one 
of the most promising models when it comes to reproduce the diphoton excess. Without any ad-hoc addition, all the
necessary ingredients are already present in the MDGSSM:
\begin{itemize}
\item There is a singlet supermultiplet $\mathbf{S}$ introduced to give the Bino a Dirac gaugino mass. It
is straightforward to identify its scalar (or pseudoscalar) component with the $750$ GeV resonance.
\item There are extra vector-like charged states, subsequently called ``fake leptons'' \cite{Benakli:2013msa} as they carry the same quantum numbers as the Standard Model leptons. They were introduced 
in order to restore the automatic gauge coupling unification that was spoiled by the addition of the adjoint representations of the Standard Model gauge group. In this work, these states will increase the coupling of the scalar resonance to photons at one loop. 
\item There is an octet supermultiplet $\mathbf{O}$ required to give the gluino a Dirac mass. This contains colour-octet scalars which will generate a coupling of the singlet resonance to gluons at one loop (via trilinear scalar couplings), required for its production in gluon fusion.
\end{itemize}

One of the important constraints to impose on any new scalar $S$ candidate to explain the excess is a bound on its mixing with the Standard Model Higgs. This mixing is not only induced at one-loop, but can be present already at tree level. The supersymmetric operator describing the Dirac gaugino bino mass leads to a modification of the $U(1)_Y$ D-term as
\begin{equation}
D_1 = D_Y^{(0)} \rightarrow D_1  =   - 2 m_{1D}  S_R +D_Y^{(0)} \qquad \textrm{with}  \qquad D_Y^{(0)}=  - g'\sum_{j} Y_j \varphi_j^* \varphi_j
\end{equation}
where $S_R$ is the real part of $S$ and $\varphi_j$ a scalar field with charge $Y_j$ under  $U(1)_Y$. Upon elimination of the auxilliary fields, this implies an interaction of the form:
\begin{equation}
g'    m_{1D}  S_R   ( |H_u^0|^2-|H_d^0|^2) .
\label{EQ:SingletDTerm}\end{equation}
thereof a tree-level induced mixing. However, this is typically compensated by the presence in the superpotential of a term of the form:
\begin{equation}
W \supset \lambda_S S H_u H_d .
\label{EQ:lambdaS}
\end{equation}
A precise evaluation of this mixing at the tree and one-loop level needs to be carried out carefully if one tries to identify the scalar $S$ in models of Dirac gauginos with a $750$ GeV
resonance. 

Our parameter space is constrained by the requirements of stability of the vacuum avoiding existence of directions in the phase space of 
the model taking the fields expectation values to charge- and
colour- breaking vacua. This is important as we shall see that trilinear terms will play an important role in generating the required amount of
scalar production and decay into di-photons. Among the trilinear terms considered here some have not been explicitly discussed in the existing literature 
while they are expected to be generically present in the model. This is the case for example of soft terms mixing three adjoint scalars that we will show that they are
 generated in models of gauge mediation.

We shall keep couplings small enough to preserve perturbativity up to the GUT scale. This restriction can be of course relaxed if one allows for Landau poles  below the GUT scale. However, as one of the virtues of the MSSM was to predict perturbative unification of gauge couplings, and was one of the motivations for introducing the MDGSSM, we shall place emphasis on finding regions of the parameter space which respect this condition. 

% We have performed a thorough study of the MDGSSM, taking into account the constraints mentioned above. We have performed many scans of different regions of the parameters that allow to reproduce the excess in the region of 750 GeV. We have found that it is quite easy to reproduce the excess if one accepts the price of some mild fine-tuning of the order of few percents which might be acceptable in view of the present bounds put by LHC on the supersymmetric extensions of the standard Model. The obtained region of parameters differs then from previous studies of the excess Dirac gaugino models \cite{Carpenter:2015ucu}, which did not consider the term \ref{notCarp}.

To find the parameter space relevant for the diphoton excess we shall use the most sophisticated tool available: the code \SARAH\cite{Staub:2008uz,Staub:2009bi,Staub:2010jh,Staub:2012pb,Staub:2013tta,Staub:2015kfa} and its \SPheno\cite{Porod:2003um,Porod:2011nf} output. This is able to calculate the masses of all particles to full one-loop order, and two-loops in the gaugeless limit for the neutral (pseudo)scalars\cite{Goodsell:2014bna,Goodsell:2015ira,Goodsell:2016udb}. It can calculate renormalisation group equations of all couplings to two-loop order, including the masses and tadpoles in Dirac gaugino models as given in \cite{Goodsell:2012fm}. A guide to its use for studying the diphoton excess was described in \cite{Staub:2016dxq}; we make some small modifications described in section \ref{SUBSEC:Numerical}. In particular, this will allow us to obtain the production and decays of our resonance at $8$ and $13$ TeV while simultaneously accurately computing its mass and assuring that the light Higgs mass is correct, and verifying that the mixing between the singlet and the Higgs is small (also computed at two loops). We shall find that quantum corrections to the spectrum of particles are not just important but essential for understanding how the model describes the excess.

Finally, we note that there have been three previous attempts to relate models with Dirac gaugino masses with the diphoton excess. In \cite{Carpenter:2015ucu} as in this work the scalar component of $S$ was the putative resonance; however, the entire coupling was driven by (\ref{EQ:SingletDTerm}) which required very large Dirac gaugino masses (which would potentially flatten the Higgs potential). In \cite{Chakraborty:2015gyj} the candidate is a neutral component of a scalar doublet $R_u^0$ introduced in the MRSSM to preserve R-symmetry, but the model required the R-symmetry to be broken to fit the excess and the Dirac nature of the gauginos played little role. As we were preparing to submit this work, \cite{Cohen:2016kuf} appeared, where the pseudoscalar component of $S$ plays the role of the resonance; it couples entirely via superpotential couplings to coloured and charged fermions and thus requires large Majorana gaugino masses and charginos close to the threshold of $375$ GeV to generate the couplings to photons and gluons. Here we will not require any Majorana masses, and will include only ingredients already allowed in the MDGSSM.

The paper is organised as follows. In section \ref{SEC:MDGSSM}, we summarise the MDGSSM field content and interactions. To generate a large gluon coupling we require trilinear scalar adjoint couplings, the generation of which we describe in section \ref{SEC:TRILINEARS} along with some observations on adjoint scalar masses. We discuss the constraints on the model in section \ref{SEC:CONSTRAINTS}; in particular, this includes a detailed study of vacuum stability, and an analysis of the constraints on colour octet scalars which are important and interesting in the context of this model. Our numerical results are provided in section \ref{SEC:EXCESS} with some benchmark points to illustrate how our model reproduces the signal. Our results are summarised in the conclusions.

%% file: ModelContent.tex
In this section we review the main ingredients of the  Minimal Dirac Gaugino Supersymmetric Standard Model (MDGSSM) introduced in~\cite{Benakli:2014cia}. 

\subsubsection*{Field content}
The MDGSSM field content can be seen as the minimal set providing the MSSM gauginos a Dirac masss while preserving  two-loop unification and perturbativity of gauge couplings. We summarised it in Table~\ref{table_fields}. In addition to the chiral multiplets transforming under the adjoint representations of the gauge groups, it includes new fields charged under the lepton number global symmetry. They consist of extra Higgs-like doublets \footnote{The hypercharges are opposite with respect to the Higgs doublet in the MSSM to match the MRSSM notation for the same fields.} $\mathbf{R}_u, \mathbf{R}_d$ as well as two pairs of vector-like right-handed electron superfields $\mathbf{{E'}}_{1,2}$ in $(\mathbf{1} ,\mathbf{1})_{1}$ and $\mathbf{\tilde{{E'}}}_{1,2}$ in $  (\mathbf{1} ,\mathbf{1})_{-1}$. Such states are compatible with an $(SU(3))^3$ Grand Unification gauge group. This is the minimal set which enables a ``natural'' unification (unification without mass thresholds tuning) similar to the MSSM. 
\begin{table}[!ht]
\begin{center}
\begin{tabular}{c|c|c|c|c|c}
\hline
\small{Names}  &                 & Spin 0                  & Spin 1/2 & Spin 1 & \scriptsize{$(SU(3), SU(2), U(1)_Y)$} \\
\hline
\rule{0pt}{2.5ex} \small{Quarks}  & $\mathbf{Q}$   & $\tilde{Q}=(\tilde{u}_L,\tilde{d}_L)$  & $(u_L,d_L)$ & & (\textbf{3}, \textbf{2}, 1/6) \\
& $\mathbf{U^c}$ & $\tilde{U}^c_L$              & $U^c_L$     & & ($\overline{\textbf{3}}$, \textbf{1}, -2/3) \\
\small{($\times 3$ families)} & $\mathbf{D^c}$ & $\tilde{D}^c_L$     & $D^c_L$     & & ($\overline{\textbf{3}}$, \textbf{1}, 1/3)  \\
\hline
\rule{0pt}{2.5ex} \small{Leptons} & $\mathbf{L}$ & ($\tilde{\nu}_{eL}$,$\tilde{e}_L$) & $(\nu_{eL},e_L)$ & & (\textbf{1}, \textbf{2}, -1/2) \\
\small{($\times 3$ families)} & $\mathbf{E^c}$ & $\tilde{E}^c$    & $E^c$          & & (\textbf{1}, \textbf{1}, 1)  \\
\hline
\rule{0pt}{2.5ex} \small{Higgs} & $\mathbf{H_u}$ & $(H_u^+ , H_u^0)$ & $(\tilde{H}_u^+ , \tilde{H}_u^0)$ & & (\textbf{1}, \textbf{2}, 1/2)  \\
& $\mathbf{H_d}$ & $(H_d^0 , H_d^-)$ & $(\tilde{H}_d^0 , \tilde{H}_d^-)$ & & (\textbf{1}, \textbf{2}, -1/2) \\
\hline
\rule{0pt}{2.5ex} \small{Gluons} & $\mathbf{W_{3\alpha}}$ & & $\lambda_{3\alpha} $                       & $g$              & (\textbf{8}, \textbf{1}, 0) \\
&   & & $  [\equiv \tilde{g}_{\alpha}]$                       &                &  \\
&  &   &      &  &   \\
W    & $\mathbf{W_{2\alpha}}$ & & $\lambda_{2\alpha} $ & $W^{\pm} , W^0$  & (\textbf{1}, \textbf{3}, 0) \\
&   & &  $ [\equiv \tilde{W}^{\pm} , \tilde{W}^{0}]$ &   &   \\
&  &   &      &  &   \\
B    & $\mathbf{W_{1\alpha}}$ & & $\lambda_{1\alpha} $                       & $B$              & (\textbf{1}, \textbf{1}, 0 ) \\
&   & & $ [\equiv \tilde{B}]$                       &                &    \\
\hline
\hline
\rule{0pt}{2.5ex} \small{DG-octet}& $\mathbf{O}$ &  $O $  & $\chi_{g} $ &  & (\textbf{8}, \textbf{1}, 0) \\
&   &    &  $  [ \equiv \tilde{g}']$ &  &  \\
&  &   &      &  &   \\
\small{DG-triplet} & $\mathbf{T}$ & $\{T^0, T^{\pm}\}$ & $\{\chi_T^0, \chi_T^{\pm}\}$ &  & (\textbf{1},\textbf{3}, 0 )\\
&   &   & $[ \equiv \{\tilde{W}'^{\pm},\tilde{W}'^{0}\}]$ &  &  \\
&  &   &      &  &   \\
\small{DG-singlet}  & $\mathbf{S}$& $S$ & $\chi_{S} $   &  & (\textbf{1}, \textbf{1}, 0 ) \\
&  &  &  $ [ \equiv \tilde{B}']$    &  &   \\
\hline
\hline
\rule{0pt}{2.5ex}  \small{Higgs-like Leptons} & $\mathbf{R_u}$  & $R_u$ & $\tilde{R}_u$ &  &  (\textbf{1}, \textbf{2}, -1/2) \\
                   & $\mathbf{R_d}$  & $R_d$ & $\tilde{R}_d$ &  &  (\textbf{1}, \textbf{2}, 1/2) \\
\hline
\small{Fake electrons} &  $ \mathbf{\hat{E}}$($\times  2$)&  $\hE$ &  $\hfE$ & & (\textbf{1}, \textbf{1},1) \\
                       &  $ \mathbf{\hat{E'}}$($\times  2$)&  $\hEp$ &  $\hfEp$ & & (\textbf{1}, \textbf{1},-1) \\
\hline
\end{tabular} 
\caption{Chiral and gauge multiplet fields in the model.}
\label{table_fields}
\end{center}
\end{table}

The adjoint chiral multiplets contain new complex adjoint scalars, $S, T$ and $O$:
% which we decompose following
\begin{align}
 & S = \frac{S_R + i S_I}{\sqrt{2}} \nn \\[0.8em] \nn
 & T = \frac{1}{2\sqrt{2}}\begin{pmatrix} T_R+iT_I & \sqrt{2}(T_{+R}+iT_{+I}) \\ \sqrt{2}(T_{-R}+iT_{-I}) & -(T_R+iT_I) \end{pmatrix}  \nn \\[0.8em] 
 & O^{(a)}=\frac{O_R^{(a)} + i O_I^{(a)}}{\sqrt{2}} \end{align}
where the $S_R, O_R^{(a)},T_R, T_{-R},T_{+R} $ f and the $S_I, O_I^{(a)},T_I, T_{-I},T_{+I} $ are real scalars and pseudo-scalars, respectively. 

\subsubsection*{Lagrangian}
\label{SUBSEC:Lagrangian}
The superpotential for these fields can be written as
\begin{align}
 W = W_{Yukawa}  + W_{DG} + W_{RV}
\end{align}
where $W_{Yukawa}$ contains the usual MSSM Yukawas part 
\begin{align}
W_{Yukawa} =& Y_u^{ij} \mathbf{U^c}_i \mathbf{Q}_j \mathbf{H}_u - Y_d^{ij} \mathbf{D^c}_i \mathbf{Q}_j \mathbf{H}_d - Y_e^{ij} \mathbf{E^c}_i \mathbf{L}_j \mathbf{H_d} 
\end{align}
$W_{DG}$ contains the a priori $R$-symmetric contributions of the non-MSSM fields\footnote{Note that our coupling $\lambda_T$ is normalised differently to \cite{belanger_dark_2009,Benakli:2011kz,Benakli:2014cia}, to match the normalisation used in \SARAH.}
\begin{align}
\label{WDG}
W_{DG} =& (\mu + \lambda_{S} \mathbf{S} ) \mathbf{H_d H_u} + \sqrt{2}\lambda_T \mathbf{H_d T H_u} \nn\\
% &+ (\mu_R + \lambda_{SR} S )R_u R_d + 2\lambda_{TR} R_u T R_d  \nn\\
% &+ (\mu_{\hE\, ij}+ \lambda_{S\hat{E}\, ij}S)\hE_i \hEp_j+ \lambda_{SEij}S E_i \hEp_j \\
% &+ \lambda_{SLRi} S L_i R_d + 2\lambda_{TLRi} L_i T R_d - Y_{\hE i} R_u H_d \hE_i\nn\\
% &  - Y_{\hEp i} R_d H_u \hEp_i  - Y_{LFV}^{ij} L_i \cdot H_d \hE_j - Y_{EFV}^{j} R_u H_d E_j \nn
&(\mu_R + \lambda_{SR} \mathbf{S} ) \mathbf{R_u R_d} + 2\lambda_{TR} \mathbf{R_u T R_d}  \nn \\
&+ (\mu_{\hE\, ij}+ \lambda_{S\hat{E^c}\, ij}\mathbf{S})\mathbf{\hE}_i \mathbf{\hEp}_j+ \lambda_{SEij}\mathbf{S} \mathbf{E^c}_i \mathbf{\hEp}_j \\
&+ \lambda_{SLRi} \mathbf{S} \mathbf{L}_i \mathbf{R}_d + 2\lambda_{TLRi} \mathbf{L}_i \mathbf{T R_d} - Y_{\hE i} \mathbf{R_u H_d} \mathbf{\hE}_i\nn\\
&  - Y_{\hEp i} \mathbf{R_d H_u} \mathbf{\hEp}_i  - Y_{LFV}^{ij} \mathbf{L_i} \cdot \mathbf{H_d} \mathbf{\hE}_j - Y_{EFV}^{j} \mathbf{R_u H_d} \mathbf{E^c}_j \nn \ ,
\end{align}
while $W_{RV}$  gathers the R-symmetry violating terms
\begin{align}
\label{WDG_RPV}
W_{RV} =& L \mathbf{S}  + \frac {\hat{M}_1}{2}\mathbf{S}^2 + \frac{\kappa}{3}
\mathbf{S}^3 + \hat{M}_2 \textrm{tr}(\mathbf{TT}) + \hat{M}_3 \textrm{tr}(\mathbf{OO}) \nn\\
&+\lambda_{ST} \mathbf{S}\textrm{tr}(\mathbf{TT}) +\lambda_{SO} \mathbf{S}\textrm{tr}(\mathbf{OO})
 + \frac{\kappa_O}{3} \textrm{tr}(\mathbf{OOO}) \nn\\
\underset{\mathrm{R-symmetry}}{\longrightarrow}& 0 \ 
\end{align}
In this work we shall, as in \cite{Benakli:2014cia}, consider scenarios where R-symmetry is preserved by the superpotential (and thus these terms vanish). However we shall also consider the possibility that they do not vanish -- so the superpotential violates R, in particular $\lambda_{SO}$ will play an important role in the following.

For simplicity and to avoid lepton-flavour-violation constraints, we shall only the terms of the first three lines of  \eqref{WDG} to appear with sizable couplings; the contributions of the last two must be small enough to be negligible for the purpose of this work, so we shall set them to zero throughout.

For the soft SUSY-breaking terms, from the MSSM we retain only the bilinear terms -- i.e. conventional mass-squared terms and the $B_\mu$ term. All the scalar triilinear and Majorana gaugino mass terms violate R-symmetry; while for $B_\mu$ we suppose that, since R-symmetry is a chiral symmetry, we are breaking R-symmetry in the Higgs sector -- and in fact it is only in combination with the superpotential terms $mu, \lambda_S, \lambda_T$  R is violated. Hence in principle we can have an entirely R-preserving supersymmetry-breaking sector. 

The soft SUSY breaking terms beyond those of the MSSM consist of\footnote{We suppress gauge indices while retaining generation indices and denote the complex conjugation of fields by upper versus lower indices.}: 
\begin{itemize}
\item Dirac gaugino masses:
\begin{align}
\label{Eq:DiracMasses}
W_{\rm{supersoft}} =& \int d^2 \theta \sqrt{2} \theta^\alpha \bigg[ m_{D1} \mathbf{S} W_{Y\,\alpha} + 2m_{D2} \tr (\mathbf{T} W_{2\,\alpha}) + 2m_{D3} \tr (\mathbf{O} W_{3\,\alpha})    \bigg].
\end{align}
\item soft terms associated with the adjoint scalars
\begin{align}
\label{eq:soft_DGscalar}
- \Delta\mathcal{L}^{\rm scalar\ soft}_{\rm adjoints} = & m_S^2  |S|^2 + \frac{1}{2} B_S
(S^2 + h.c.)  + 2 m_T^2 \textrm{tr}(T^\dagger T) + (B_T \textrm{tr}(T T)+ h.c.) \nonumber \\   
&+ 2 m_O^2 \textrm{tr}(O^\dagger O) + (B_O \textrm{tr}(OO)+ h.c.) \nn \\ 
&+ \big[T_S SH_u\cdot H_d +  2 T_T H_d \cdot T H_u + \frac{1}{3} \kappa  A_{\kappa} S^3 + t_S S + h.c.\big] \nn\\
&+ \big[ T_{SO} S \mathrm{tr} (O^2) + T_{ST} S \mathrm{tr} (T^2) + \frac{1}{3} T_O \mathrm{tr} (O^3) + h.c. \big] 
%\label{Lsoft-DGAdjoint}
\end{align}
The terms on the last line have generally been neglected, but will play an important role in this work.
\item  soft terms involving the new vector-like leptons:
\begin{align}
\label{softfake}
- \Delta\mathcal{L}^{\rm scalar\ soft}_{\rm vector-like} =& m_{R_u}^2 |R_u|^2 + m_{R_d}^2 |R_d|^2 + [ B_R R_d R_u + h.c.] \nn\\
& + \hE_i (m_{\hE}^2)^i_j \hE^j +\hEp{}^i (m_{\hEp}^2)_i^j \hEp_j + [ B_{\hE}^{ij} \hE_i \hEp_j + h.c.]  \nn\\
& + [ T_{SE}^{ij} S \hE_i \hEp_j +  T_{SR} S R_d R_u+ h.c. ] \ .
\end{align}
\end{itemize}

% Overall, we have an important number of parameters. In this work, we will consider that the couplings between the fake and true leptons, which are given in the two last lines of~\eqref{WDG} are negligible. Notice that there are important constraints on them from rare leptons decay and electron dipole moment \luc{*** Stupid question, can these operators be used for the anomalous muon Muon g−2 ? ***}.\cite{benakli_constrained_2014} Furthermore, we will suppose that the terms in the last lines of~\eqref{eq_soft_DGscalar} are small so that we will not study their phenomenology.

Let us highlight that in an R-symmetry conserving model, one cannot simultaneously have the trilinears $T_{SE}$ (respectively $T_{SR}$) from~\eqref{softfake} and the superpotential couplings $\lambda_{SE}$ (respectively $\lambda_{SR}$) from~\eqref{WDG} as each term requires a different R-charge for the fields $\hE$ and $\hEp$ (respectively $R_u$ and $R_d$) to be R-invariant.

\subsubsection*{Scalar mass matrix}

We use the notation
\begin{eqnarray}
\tilde{m}_{S}^2&=&\tilde{m}_{SR}^2
+\lambda_S^2 \,  \frac{v^2}{2}  \nn\\
\tilde{m}_T^2 &=& \tilde{m}_{TR}^2 +\lambda_T^2  \frac{v^2}{2} \ ,
 \end{eqnarray}
where the effective masses for the real parts of  $S$ and $T$ read:
\begin{eqnarray}
\tilde{m}^2_{SR}& = & m_S^2+4 m^2_{1D}+ B_S, \qquad  \, \tilde{m}^2_{TR}= 
m_T^2+4 m^2_{2D}+ B_T \ .
 \end{eqnarray}
Then, at tree level the scalar mass matrix in the basis $\{h, H,
S_R,T^0_R\}$ is~\cite{Benakli:2012cy}:
\begin{eqnarray}
\label{eq:MassScalar}
\left(\begin{array}{c c c c }
M_Z^2+\Delta_h s_{2\beta}^2 & \Delta_h s_{2\beta}  c_{2\beta}  & \Delta_{hS}    
& 
\Delta_{hT} \\
\Delta_h s_{2\beta}  c_{2\beta} & M_A^2-\Delta_h s_{2\beta}^2   & \Delta_{HS}    
&\Delta_{HT}    \\
 \Delta_{hS}     & \Delta_{HS}      & \tilde{m}_S^2  &  \lambda_S \lambda_T
\frac{v^2}{2} \\
  \Delta_{hT}     &\Delta_{HT}    &  \lambda_S \lambda_T \frac{v^2}{2} & 
\tilde{m}_T^2  \\
\end{array}\right) 
\end{eqnarray}
where we have defined:
\begin{eqnarray}
\Delta_h&=&\frac{v^2}{2}(\lambda_S^2+\lambda_T^2)-M_Z^2 
\end{eqnarray}
which vanishes when $\lambda_S$ and $\lambda_T$ take their $N=2$ values, 
\begin{eqnarray}
\label{mixinghs}
 \Delta_{hS} =- 2 \frac {v_ S} {v}   \tilde{m}_{SR}^2, \qquad  
 \Delta_{hT} = -2 \frac {v_ T} {v}   \tilde{m}_{TR}^2
\end{eqnarray}
and
\begin{eqnarray}
 \Delta_{HS} = g' m_{1D} v  s_{2\beta} ,\qquad
 \Delta_{HT}  =  - g m_{2D} v  s_{2\beta}  \nonumber \\ \ .
 \end{eqnarray}
This matrix is diagonalised by the mixing matrix $S_{ij}$. Of particular interest will be $S_{11}$ which measures if the lightest scalar eigenstate is Standard Model Higgs like, and $S_{13}$ which measures the proportion of the scalar singlet $S_R$ in this lightest eigenstate.

%% file: Trilinears.tex
Previous studies of Dirac gaugino models have generally neglected the phenomenology of adjoint self-coupling terms, with an exception being a superpotential term $\frac{\kappa}{3} S^3$ used in \cite{Benakli:2011kz} to generate $\mu/B_\mu$ as in the NMSSM, and a recent brief discussion in \cite{Carpenter:2015mna}. In the case of superpotential terms such as $\lambda_{SO}$ these can be neglected when considering an R-symmetric visible sector; however, trilinear soft couplings such as $T_{SO}, T_O$ (see (\ref{eq:soft_DGscalar})) are always allowed. It is therefore interesting to consider what values we expect from models of supersymmetry-breaking mediation. 

Starting with a spurion analysis where supersymmetry is broken by either a D-term $D$ or F-term $F$, then if the mediating dynamics is at a scale $M$ the terms in our effective Lagrangian should be given by powers of $\frac{D}{M}, \frac{F}{M}, \frac{D}{M^2}, \frac{F}{M^2}$ with appropriate factors of couplings and $\kappa_l \equiv 1/16\pi^2$. Furthermore, quartic and higher-order couplings -- which are ``hard'' SUSY-breaking parameters -- are always generated, but do not lead to quadratic divergences because they appear suppressed by powers of the scale $M$ which is the cutoff of our effective theory. Important in this work are the quartics such as $\mathcal{L} \supset \frac{\lambda_{4S}}{24} S^4  $ which must have size $\lambda_{4S} \sim \kappa_l^p \left(\frac{D}{M^2} \right)^q $ for some integer $p, q$ (or similarly for F-terms with \emph{even} $q$)); taking $p=1,q=1$ for a D-term we naively have a quadratic divergence in the scalar mass proportional to $\lambda_{4S}$ but this yields $\Delta m^2_S \sim \kappa_l \lambda_{4S} M^2 \sim \kappa_l^2 D \ll M^2$, while for $q=2$ we have $\kappa_l^2 \frac{D^2}{M^2} $. In fact, this tells us that the case $q=1$ is special because it implies a much larger correction at one loop than the direct mass, and could therefore destabilise the calculation. We shall return to this below.

As a first observation, if the mediation is by gravity, then $M$ should be identified with the Planck scale (unless there is significant sequestering) and we should only consider the leading order terms. We would therefore require the quantum gravity theory to give us the terms $T_{SO}, T_O$ at leading order $D/M, F/M$ and the quartics must, by the above reasoning, be negligible. 

On the other hand, in the case of low-scale supersymmetry breaking -- where it was argued in \cite{Gherghetta:2011na} that this requires Dirac gauginos -- $\sqrt{F} \sim \sqrt{D} \sim M \sim$ TeV, and we generate all terms at a similar order, which would include $T_{SO}, T_O$. However, the phenomenology is significantly changed by the presence of higher-dimensional operators and the goldstino couplings \cite{Goodsell:2014dia} and, since it is difficult to reconcile with perturbative unification, we shall not discuss this further here. 

Finally, for gauge mediation $M$ could be as small as $\sqrt{F}$ or $\sqrt{D}$ but there is no a priori upper limit on $M$ until we choose a particular quantum gravity embedding. The Dirac gaugino masses are expected to be generated at one loop and be of order $\kappa_l \frac{D}{M}$ or $\kappa_l \frac{F^2}{M^3}$. For F-term breaking the standard gauge-mediation soft mass-squareds for the squarks/sleptons are of order $\kappa_l^2 \frac{F^2}{M^2}$, while in D-term breaking they may be suppressed. Therefore if we imagine that $\kappa_l \frac{D}{M} \sim $ TeV, then for terms $\kappa_l \frac{D^2}{M^3} $ to be significant we would need $D \sim M^2$ and furthermore $M \sim 100$ TeV. %Ideally, to avoid this constraint, we should require all important operators to be generated at leading order in $D/M^2, F/M^2$. 

\subsection{Adjoint couplings in gauge mediation}

One of the most interesting issues in the construction of gauge mediation models with Dirac gaugino masses has been that of the adjoint scalar masses: in the simplest realisation, only a B-type mass-squared $\mathcal{L} \supset - \frac{1}{2} B_\Sigma \Sigma^2$ is generated at leading order in $D/M^2$, and not a conventional mass-squared $ \mathcal{L} \supset - m_\Sigma^2 |\Sigma|^2$. This happens for one pair of vector-like messengers $Q, \tilde{Q}$ having charges under a hidden $U(1)$ of $+1,-1$, where the $U(1)$ obtains a D-term. This was noticed from the earliest models \cite{Fox:2002bu,Antoniadis:2006uj} with the original proposed solution being to add a supersymmetric mass for the adjoint -- which would also violate the R-symmetry and generate Majorana masses for the gauginos, with a see-saw effect. However, an alternative solution was found to be to introduce additional messenger states with non-diagonal couplings to either the adjoints (in the D-term case) \cite{Benakli:2008pg,Benakli:2010gi} or an F-term spurion \cite{Benakli:2008pg,Amigo:2008rc,Benakli:2010gi}; in the D-term case this requires the couplings to violate the $U(1)$-charges. In \cite{Benakli:2010gi} examples were given where the ratio of B-type to conventional masses is arbitrary. The general ansatz was to couple the adjoint to messenger fields $Q_i, \tilde{Q}_j$ and to possible F-term spurions $X$ via superpotential couplings
\begin{align}
W \supset M Q_i \tilde{Q}_i + \lambda_{i\tilde{j}} Q_i \Sigma \tilde{Q}_j + \mu_{i\tilde{j}} X Q_i \tilde{Q}_j
\label{EQ:BGansatz}\end{align}
and D-terms via charges $e_i, \tilde{e}_i$ which we can write as a matrix $e_{i \ov{j}} (Q_i Q_j^* -\tilde{Q}_i \tilde{Q}_j^*) $.  

 More recently, the issue has been re-examined. One suggested approach, dubbed ``Goldstone gauginos,'' is to promote the adjoints to be the Goldstone bosons of a broken symmetry\cite{Alves:2015kia,Alves:2015bba}; however, this solution would lead effectively to no higher-order interactions for our adjoint scalars and we do not consider it here. More in the spirit of the earlier works, the issue was rephrased in the language of effective operators in \cite{Carpenter:2010as,Csaki:2013fla,Carpenter:2015mna}, where it was claimed that the explanation for the absence of conventional mass-squared terms for the adjoints at leading order is that the operator responsible for the generation of a leading-order mass-squared term should be 
\begin{align}
\mathcal{L} \supset \int d^4 \theta \frac{1}{M^2} [ \psi^\dagger e^{qV} \psi + \tilde{\psi}^\dagger e^{-qV} \tilde{\psi} ] \Sigma^\dagger \Sigma,
\end{align}
where $\psi, \tilde{\psi}$ are a pair of fields charged under the hidden $U(1)$ with charges $\pm q$ which obtain vevs (and thus generate a contribution to the hidden D-term). The above operator is generated by including terms in the superpotential that mix the messengers $Q, \tilde{Q}$ with other pairs of fields $N, \tilde{N}$ which are neutral (or at least have different charges) under the hidden $U(1)$, so that the vevs of $\psi, \tilde{\psi}$ generate messenger mixing terms. This is clearly nearly equivalent to the above ansatz, and can be written in the form
\begin{align}
W \supset M_{ij} Q_i \tilde{Q}_j + \lambda_{i} Q_i \Sigma \tilde{Q}_i
\label{EQ:MixingAnsatz}\end{align}
where we now write the mass terms as violating the $U(1)$ charges instead. 

If we start with the case of no couplings/mass mixing terms that violate the $U(1)$ D-term charges, we shall first give a simple proof that the conventional mass term $|\Sigma|^2$ vanishes at leading order for any number of messengers, and then look at higher-order terms. Considering first the visible gauge group to be $U(1)$, we have the effective potential contribution from the messenger scalars (since the fermion potential is independent of $D$):
\begin{align}
V =& \int \frac{d^d q}{(2\pi)^d}\mathrm{tr} \log (q^2 + \mathcal{M}_Q^2 + De) + \mathrm{tr} \log (q^2 + \mathcal{M}_{\tilde{Q}}^2 - De)\nn\\
\equiv& V_+ + V_- .
\end{align}
Here we have $\mathcal{M}_Q^2 = (M + \lambda \Sigma)(M^\dagger + \lambda^\dagger \ov{\Sigma})$, $\mathcal{M}_{\tilde{Q}}^2 = (M^\dagger + \lambda^\dagger \ov{\Sigma})(M + \lambda \Sigma)(M^\dagger + \lambda^\dagger \ov{\Sigma})$ are the supersymmetric mass-squared matrices.
Now, if we take the couplings to preserve the $U(1)$ charges then we can write the 
\begin{align}
V_+=& De\int \frac{d^d q}{(2\pi)^d}\mathrm{tr} \left( \frac{1}{q^2 + \mathcal{M}_{Q}^2} \right)  - \frac{1}{2} D^2e^2\int \frac{d^d q}{(2\pi)^d}\mathrm{tr} \left( \frac{1}{q^2 + \mathcal{M}^2_{Q}} \right)^2 + \mathcal{O}(D^3) \nn\\
\rightarrow 16\pi^2 V =& D^2e^2 \mathrm{tr} \bigg(\log \mathcal{M}_{Q}^2/\mu^2  \bigg)+ \mathcal{O}(D^4)
\end{align}
since the eigenvalues of $\mathcal{M}_{Q}^2$ and $\mathcal{M}_{\tilde{Q}}^2$ are equal. Next, by taking the derivative with respect to $\Sigma$ we find only a holomorphic function of $\Sigma$: 
\begin{align}
16\pi^2 \frac{\partial V}{\partial \Sigma} =& D^2 e^2 \mathrm{tr} \bigg( [ M + \lambda \Sigma]^{-1} \lambda \bigg) + \mathcal{O}(D^4) \nn\\
\rightarrow V =& \frac{D^2e^2}{16\pi^2}\bigg[  \mathrm{tr}\bigg(\log M M^\dagger/\mu^2  \bigg) + \tilde{V} (\Sigma) + \ov{\tilde{V}} (\ov{\Sigma}) \bigg] + \mathcal{O}(D^4).
\end{align}
As an example, consider the simple model of a single messenger where the matrices become numbers; then we have
\begin{align}
\tilde{V} (\Sigma) =& -\sum_{n=1}^\infty \frac{1}{n}\left( \frac{-\lambda \Sigma}{M} \right)^n. 
\label{EQ:SimpleModelPot}\end{align}
This potential manifestly has trilinear and quartic couplings, although at order $\frac{D^2}{M^3}, \frac{D^2}{M^4}$ respectively. Indeed, if we continue with the ansatz (\ref{EQ:BGansatz}) then it is easy to see that there are no terms of linear order in $D$, because $\mathcal{M}_Q^2 = (M+\lambda \Sigma)(M+ \lambda^\dagger \Sigma^\dagger) = \mathcal{M}_{\tilde{Q}}^2$ and
\begin{align}
V =& D\int \frac{d^d q}{(2\pi)^d}\mathrm{tr} \bigg\{ \bigg([q^2 + \mathcal{M}_Q^2]^{-1} - [q^2 + \mathcal{M}_{\tilde{Q}}^2]^{-1} \bigg) e \bigg\} + \mathcal{O}(D^2).
\end{align}
Hence to have large cubic interactions we should start from ansatz (\ref{EQ:MixingAnsatz}). In this way, in order to have an interesting phenomenology we require either $D \sim M^2$ with both at a low scale, or we require (as proposed in \cite{Csaki:2013fla}) that
\begin{align}
B_\Sigma < m_\Sigma^2 \sim a D + b \frac{D^2}{M^2} 
\end{align}
with some cancellation between the two terms so that we can have $m_\Sigma \sim T_{SO}$. Note that once we take this ansatz with non-zero mixing between the messengers and $[M, e] \ne 0$ we typically generate trilinear terms in the potential -- but also tadpoles. The issue of tadpoles is then easily circumvented by embedding the coupling of the singlet adjoint $S$ to the $SU(3)$ and $SU(2)$ adjoints into the generator $T^Y = \frac{1}{\sqrt{60}}\mathrm{diag}(2,2,2,-3,-3)$. This then also means that the couplings of the singlet adjoint $S$ are related to those of $T$ and $O$; for example, for $T_{SO}$, if we have calculated the coupling for $U(1)$ messengers as being $\mathcal{L}\supset \frac{1}{6} T_\Sigma \Sigma^3$, then we have
\begin{align}
T_{SO} \delta^{ab}=& T_\Sigma \mathrm{tr} ( T^Y T^a_3 T^b_3)  \nn\\
=& \frac{1}{\sqrt{15}} T_\Sigma \delta^{ab} 
\end{align}
where $T^a_3, T^b_3$ are $SU(3)$ generators. However, exploring sets of messengers which give these desired properties with sufficiently large trilinear couplings and exploring the vacuum stability of the total system would be very interesting, but is beyond the scope of this work.

%% file: productionMDGSSM.tex
In the narrow width approximation in which the mediating $S$ singlet is automatically on-shell, we can approximate the cross section of the complete process $pp \rightarrow S \rightarrow \gamma \gamma$ as follows:
\begin{align}
\sigma(pp \rightarrow S \rightarrow \gamma \gamma) =~& \frac{2J+1}{s \ms \Gamma} \bigg[ C_{gg} \Gamma(S \rightarrow gg) + \sum_q C_{q\ov{q}} \Gamma (S \rightarrow q \ov{q}) \bigg] \Gamma(S \rightarrow \gamma \gamma)\,.
\end{align}
Assuming a spin-zero particle produced resonantly via gluon fusion, we arrive at 
\begin{align}
\label{prodrate}
\sigma(pp \rightarrow S \rightarrow \gamma \gamma)_{13\TeV} \approx~ & K_{13} \times 4.9 \times 10^{6} \fb \frac{\Gamma_{gg}}{\Gamma} \frac{\Gamma_{\gamma \gamma}}{\Gamma} \frac{\Gamma}{\ms} \\
\sigma(pp \rightarrow S \rightarrow \gamma \gamma)_{8\TeV} \approx~ & K_8 \times 1.1 \times 10^{6} \fb\frac{\Gamma_{gg}}{\Gamma} \frac{\Gamma_{\gamma \gamma}}{\Gamma} \frac{\Gamma}{\ms} \nn \ ,
\end{align}
taking $C_{gg}^{8 \mathrm{TeV}} = 174$ and $C_{gg}^{13 \mathrm{TeV}} = 2137$ as values arising from the parton distribution functions \cite{Franceschini:2015kwy}, respectively. An important aspect of our calculation is that for a more realistic estimation, we have taken into account the K-factors $K_{8,13}$ for the full N$^n$LO production of $H+\mathrm{jet}$ compared to the tree-level process. We have estimated $K_8 \simeq 1.9$ from the comparison of the leading-order effective vertex from {\tt MadGraph} and the Higgs Cross-section working group value for a Standard-Model-like Higgs of $750$ GeV at $8$ TeV. We will take conservatively the same value for $K_{13}$.

% The two relevant amplitudes $\G_{gg}$ and $\G_{\g \g}$ corresponding to the loop processes $\Stogg$ and $\Stogluglu$ are obtained from one-loop diagrams of the form of Figure~\ref{fig:loopCouplingS}.
%  \begin{figure}[t]
%  \begin{center}
%  \includegraphics[width=0.65\textwidth]{\pdir/diphoton.png}
%  \caption{ One-loop diagrams contributing to the couplings of $S$ into two gluons and two photons.}
%  \vspace*{-2mm}
%  \label{fig:loopCouplingS}
%  \end{center}
%  \end{figure}

Let us first consider the coupling to two gluons. The process $\Stogluglu$ is a priori generated by loops of squarks, scalar octet and gluinos. The amplitude is of the form 
\begin{align}
\Gamma (S \rightarrow g g) ~=& ~\frac{\alpha_3^2 m_S}{8 \pi^3} \left| \tr \left( \sum_f  C_f \frac{g_{Sff} }{\sqrt{\tau_f}} A_{1/2}^S (\tau_f) + \sum_\phi C_\phi \frac{g_{S\phi\phi}}{2\sqrt{\tau_\phi} m_\phi}  A_0^S (\tau_\phi) \right)\right|^2 \\
~\simeq & ~ 4.3 \cdot 10^{-2} \left| \tr \left( \sum \frac{g_{Sff} }{\sqrt{\tau_f}}A_{1/2}^S (\tau_f) + \frac{g_{S\phi\phi}}{2\sqrt{\tau_\phi} m_\phi} A_0^S (\tau_\phi) \right)\right|^2 \nn \ ,
\end{align}
where we have defined $ \tau_i \equiv  4 \frac{m_i^2}{m_S^2}  $, the sums runs over all scalars and fermions, and
\begin{align}
f(\tau) \equiv~& \left\{ \begin{array}{cl} (\sin^{-1} (1/\sqrt{\tau}))^2 & \tau \ge 1 \\ -\frac{1}{4} \bigg[ \log \frac{1+\sqrt{1-\tau}}{1-\sqrt{1-\tau}} - i\pi \bigg]^2 & \tau < 1 \end{array} \right. \nn\\
A_0^S =~& \tau ( \tau f(\tau) - 1) \nn\\
A_{1/2}^S =~& 2\tau \big( 1+ (1-\tau)f(\tau))\big) \nn \ .
\end{align}
and $Q_f$, $Q_\phi$, $g_{Sff}$ and $g_{S\phi\phi}$ are the electric charge and coupling with the singlet of the fermions and scalars participating in the triangular loops. The loop fonctions $A_0^S$ and $A_{1/2}^S$ have a maximum at the resonant mass $\ms / 2 \sim 375 $ GeV. We will therefore generically require masses close to this scale in order to enhance the cross-section. The main contributions to the loop will be:
\begin{itemize}
 \item D-term-induced couplings between the squarks and the singlet, generated by the Dirac masses operator of Eq.~\eqref{Eq:DiracMasses}. Theses couplings are proportional to the hypercharge of the squarks and the Dirac mass $m_{1D}$. They are sizeable only for large Dirac mass $m_{1D}$.
\item Soft terms trilinears couplings from~\eqref{softfake} between the adjoint scalar octet and the singlet. They give a sizeable contribution but unfortunately are strongly constrained from vacuum stability bounds.
\end{itemize}
A priori, one could have expected a contribution from the Dirac gluinos. However, we observed that pure Dirac gluinos do not contribute at all to the amplitude. This remark is of crucial importance for the pseudo-scalar $S_I$ which can only couple to gluons through fermions loops as we assume CP-conserving interactions. If no Majorana masses for the original gluinos are introduced, the pseudo-scalar is practically not produced. However, if we allow for the presence of an additional Majorana mass term, the pseudo-scalar $S_I$ can then participates in the $\Stogg$ cross-section, potentially leading to a ``double-peaks'' scenario, as we will see later.

We now turn to the amplitude to diphoton. This is given for a scalar by
\begin{align}
\label{AmplitudeGamma}
\Gamma (S \rightarrow \gamma \gamma) ~=& ~\frac{\alpha^2 m_S}{64 \pi^3} \left| \tr ( \sum_f  \frac{g_{Sff} }{\sqrt{\tau_f}} Q_f^2 A_{1/2}^S (\tau_f) + \sum_{\phi} \frac{g_{S\phi\phi}}{2\sqrt{\tau_\phi} m_\phi} Q_\phi^2 A_0^S (\tau_\phi) )\right|^2 \\
 ~\simeq & ~ 2.0\cdot 10^{-5} \left| \tr ( \sum \frac{g_{Sff} }{\sqrt{\tau_f}}A_{1/2}^S (\tau_f) + \frac{g_{S\phi\phi}}{2\sqrt{\tau_\phi} m_\phi} A_0^S (\tau_\phi) )\right|^2 \nn \ .
\end{align}
In order to get an idea of the enhancement we need from the square term, let us find the smallest value of $\G_{\g\g}$ leading to a $\sigma(\Stogg) \gtrsim 2$ fb. In the limit in which $\Gamma_{gg}$ dominates the decay width, we can use Eq.~\eqref{prodrate} to get 
\begin{align}
\label{boundGgg}
\G_{\g\g} \gtrsim  1.6 \times 10^{-4} ,
\end{align}
which is an order of magnitude bigger than the numerical factor in~\eqref{AmplitudeGamma}. The key issue will therefore be to populate the sums in the square terms of~\eqref{AmplitudeGamma} since the amplitude will very roughly scale as $N^2$ , with $N$ the number of particles participating in the loop. The main contributions will come from
\begin{itemize}
 \item D-term-induced couplings between the sleptons and the singlet, they are again proportional to the hypercharge of the sleptons and to the Dirac mass $m_{1D}$. They are therefore sizeable only for large Dirac mass.
 \item Superpotential-induced couplings between the fake leptons and the singlet from the terms of~\eqref{WDG} in section~\ref{SEC:MDGSSM}. They are the main contributions in our model.\footnote{Notice that since the coupling $\lambda_S$ is usually small in most of the scenarios we will consider, the Higgsinos contribution will also be small.}
\item Soft terms trilinears couplings from~\eqref{softfake} between the fake sleptons and the singlet. They are again strongly constrained from vacuum stability bounds.
\end{itemize}
An important remark here is that the two last contributions are mutually incompatible in presence of a preserved R-symmetry as we already stressed in Section~\ref{SUBSEC:Lagrangian}.

% Finally, a theoretical estimation for the cross-section in the ideal case where the gluon amplitude dominate the decay width and there is no mixing between the scalar singlet and the Higgs boson is given in Figure~\ref{fig:Amplitudes}. 
% \begin{figure}[!htbp]
% \begin{center}
% \includegraphics[height=0.4\textheight]{Plots/Amplitudes_msl.jpg}
% \caption{Cross-section of  $S \rightarrow \gamma \gamma$  in fb as a function of the pole mass of the right-handed slepton $m_{\tilde{l}_R}$ and of the fake lepton $\mu_E$. We set the fake sleptons trilinears to zero and assume heavy fake slepton. We assume that the gluon fusion process dominates the other processes. We have taken $\lambda_{SR} = \lambda_{ER} = 0.7$.}
% \label{fig:Amplitudes}
% \end{center}
% \end{figure}

%% file: Mixing.tex
A crucial property of the singlet $S$ is that it will in general mix with the Higgs eigenstates. This is in our case an undesirable feature since it will lead to tree-level decays of $S$ into tops, $W$, $Z$ or Higgs which could easily overcome the one-loop decay into photons.

\subsubsection*{Analytical Estimate}

Building on the notations introduced in the previous sections, we can use the minimisation condition of $v_S$ on the off-diagonal element $\Delta_{hS}$ of the scalar mass matrix given in~\eqref{mixinghs} to find (see~\cite{goodsell_two-loop_2013})
\begin{align}
\Delta_{hS} =~& v  [v_S\lambda_S^2  - g' m_{1D} c_{2\beta} + \sqrt{2} \lambda_S \mu + \lambda_S \lambda_T v_T] \nn\\
=~& v[ \sqrt{2} \lambda_S \tilde{\mu} - g' m_{1D} c_{2\beta}] \ ,
\label{EQ:HSRestriction}
\end{align}
where we used the effective mass parameter
\begin{align}
\label{defmueff}
\tilde{\mu}  =~ & \mu +   \frac{1}{\sqrt{2}} (\lambda_S \,   v_S + \lambda_T  \,  v_T) \ .
\end{align}
From this basic analytical calculation, we see that we can minimise the tree-level mixing by choosing:
\begin{align}
 \lambda_S \sim \frac{\gp m_{1D}  c_{2\beta}}{\sqrt{2} \tilde{\mu}} \ .
\end{align}
In general, this relation will be modified at one-loop, but the property that one value of $\lambda_S$ is favored will remain and is easily observable in our coming Figures.

\subsubsection*{Experimental Bounds and Naturalness}

\begin{figure}[!htbp]
\begin{center}
\includegraphics[width=0.95\textwidth]{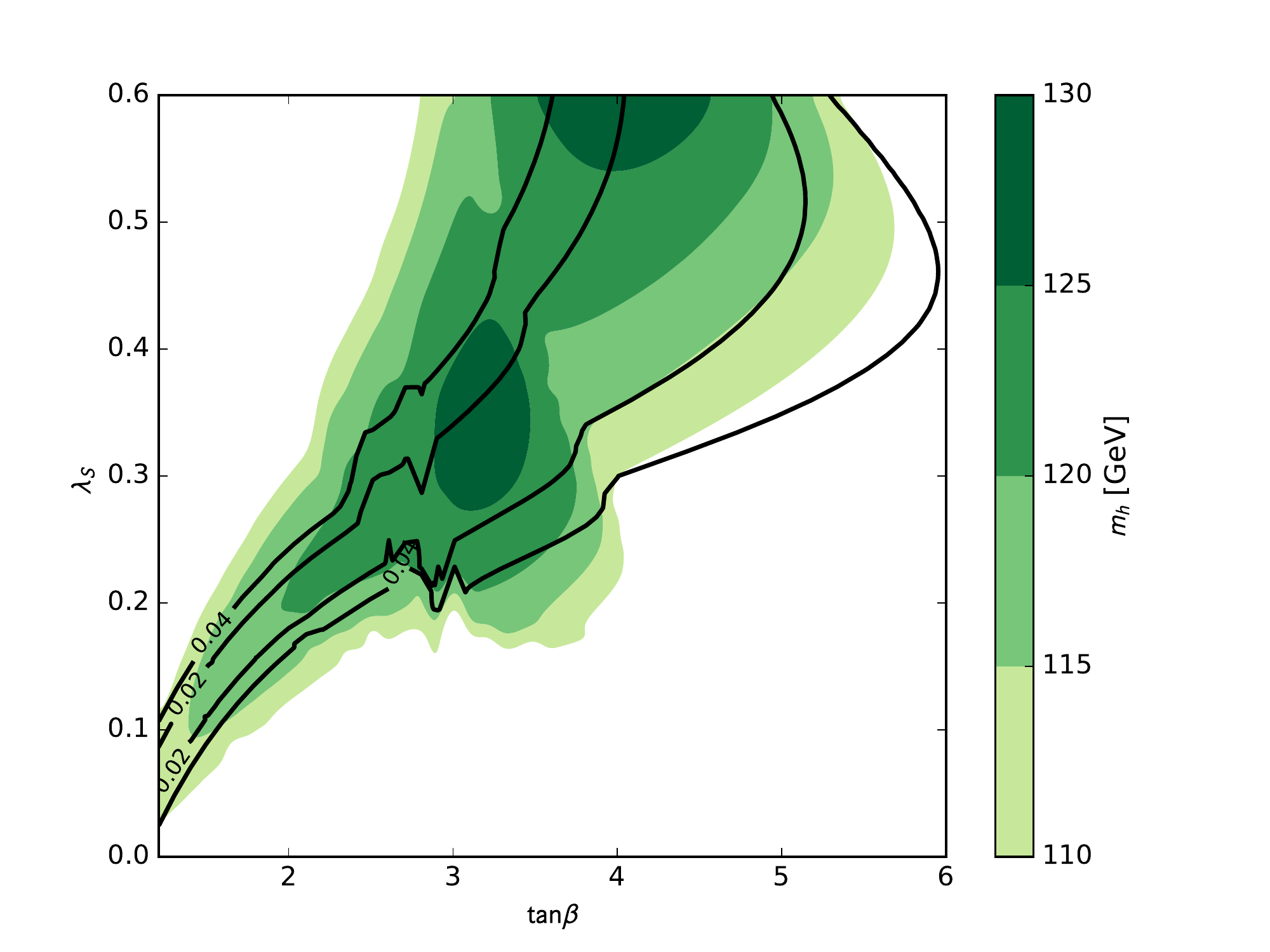}
\caption{ Higgs mass and mixing between $h$ and $S$ as a function of $\lambda_S$ and $\tan \beta$. The thin black lines represent the $2\%$ and $4 \%$ mixing contour lines. The anomalies around $\tan \beta \sim 2.5$ corresponds to the region where the two-loop effective potential used to determined the Higgs mass suffers from the so-called ``Goldstone boson catastrophy'' (see~\cite{Goodsell:2015ira} for more details). }
\label{fig:mH_tanls}
\end{center}
\end{figure}

Such a mixing with the Standard Model Higgs will modify the Higgs sector observables. From~\cite{HiggsATLASCMS} we find the latest constraint on the $125$ GeV Higgs global signal strength  $\mu_{\textrm{average}}$  to be
\begin{align}
\mu_{\textrm{average}} = 1.09^{+0.11}_{-0.10} \ ,
\end{align}
In our case this is modified by a factor of $|S_{11}|^2$, where $S$ is the mixing matrices of the scalar sector; the above constraint gives us
\begin{align}
1-|S_{11}|^2 \le 0.24 \leftrightarrow \sum_{k \ne 1} |S_{1k}|^2 = \sum_{k \ne 1} |S_{k1}|^2 \le 0.24.
\end{align}
This condition is in fact satisfied quite easily, as can be seen from Figure~\ref{fig:mH_tanls} where we show the contours for the Higgs mass and the mixing matrix element $S_{31}$ as a function of $\tan \beta$ and $\lambda_S$. An important comment regarding this Figure is that a $125$ GeV also favors small mixing.

More stringent constraints arises from the non-observation of any excess in the $8$ TeV data for the $ZZ$, and $hh$, dijets and $WW$  channels. As the mixing between $S$ and $h$ induces a tree-level decay one naively expect a percent-level suppression to be necessary. Since the $S$ is mostly produced by gluons fusions in our scenarios, we request that (see~\cite{Franceschini:2015kwy}): 
\begin{align}
\label{RatiosLHC}
 \frac{\G(S \rightarrow ZZ)}{\G(S \rightarrow \g \g)} \lesssim 6 \\ \nn
 \frac{\G(S \rightarrow Zh)}{\G(S \rightarrow \g \g)} \lesssim 10 \\ \nn
 \frac{\G(S \rightarrow hh)}{\G(S \rightarrow \g \g)} \lesssim 20
\end{align}
which gives the most stringent constraints on the mixing between $S$ and $h$. 
% Imposing these constraints further limits the possible range of values for $\lambda_S$ (for fixed values of the other parameters), as can be seen in Figure~\ref{fig:sig_ls.pdf}.
% \begin{figure}[!htbp]
% \begin{center}
% \includegraphics[height=0.38\textheight]{Plots/sig_ls.pdf}
% \caption{ $\Stogg$ cross section in $\fb$ as a function of $\lambda_S$. We have represented in red the ratio $\frac{\G(S \rightarrow ZZ)}{\G(S \rightarrow \g \g)} $ and in blue the ratio $\frac{\G(S \rightarrow hh)}{\G(S \rightarrow \g \g)}$  ** To be replaced by random scan cross-section vs higgs mass**   }
% \label{fig:sig_ls.pdf}
% \end{center}
% \end{figure}
The di-Higgs channel is proportional to the tree-level mixing term without passing through the mixing, because the vertex is given by $\Delta_{hs}/v$ (plus smaller terms proportional to the mixing matrix elements $S_{13}, S_{31}$); we have 
\begin{align}
\frac{\Gamma (s_R \rightarrow h h)}{m_{SR}}  \simeq& \left(\frac{\Delta_{hs}}{v}\right)^2 \frac{1}{32\pi m_{SR}^2} \sqrt{1 - \frac{4 m_h^2}{m_{SR}^2}} \nn\\
\simeq~&  0.01\times \left(\frac{m_{SR}^2}{v^2} \right)\left(\frac{\Delta_{hs}}{m_{SR}^2}\right)^2 \nn\\
\simeq~& 0.1 \times |S_{13}|^2,
\end{align}
which gives a constraint of $S_{13} \lesssim 0.01$. On the other hand the constraints for $Z$ and $W$ decays come purely through the mixing matrix; defining $x \equiv  \frac{m_V^2}{m_S^2} $ for a vector boson $V$ we have a decay rate
\begin{align}
\Gamma (S\rightarrow VV) =~& \frac{|c_{sVV}|^2}{128\pi m_V} x^{-3/2} ( 1 - 4 x + 12 x^2) \sqrt{1 - 4x}  
\end{align}
and
\begin{align}
c_{hZZ} =~& \frac{g_Y^2 + g_2^2}{2} v = \frac{2M_Z^2}{v}, \quad c_{hW^+W^-} = \frac{g_2^2 v}{2}   \simeq \frac{2 M_W^2}{v},  \quad c_{tW^+W^-} = 2 g_2^2 v_T.
\end{align}
Neglecting $v_T$ and mixing with the triplet as small effects, we can then write
\begin{align}
\frac{\Gamma (S\rightarrow ZZ)}{m_S} \simeq~ &  0.09 |S_{13}|^2 \nn\\
\frac{\Gamma (S\rightarrow WW)}{m_S} \simeq~ & 0.17 |S_{13}|^2.
\end{align}
Translating these into constraints, we see that it is the $Z$ decays which are most important. 

Notice that the only loop decays included in this paper are $\Stogg$ and $\Stogluglu$ (as they do not have a tree-level contribution). A priori in the negligible mixing region, one should also consider the other diboson loop decays (in particular to $Z\gamma$). However, almost all of the new fields contributing to the loop decays will be $SU(2)$ singlets so that the decay to diphoton will be the dominant diboson decay channel. The only exceptions are the new doublets $\mathbf{R}_u$ and $\mathbf{R}_d$ which should mostly decay to $WW$, $ZZ$ and $Z \gamma$. Due to the interference with the tree-level processes the loop contribution to these processes is not currently calculated in \SARAH; their implementation is eagerly awaited in future work, but here we note that they will not have a significant impact on our results as described in \cite{Staub:2016dxq}.

\begin{figure}[!htbp]
\begin{center}
\includegraphics[width=0.99\textwidth]{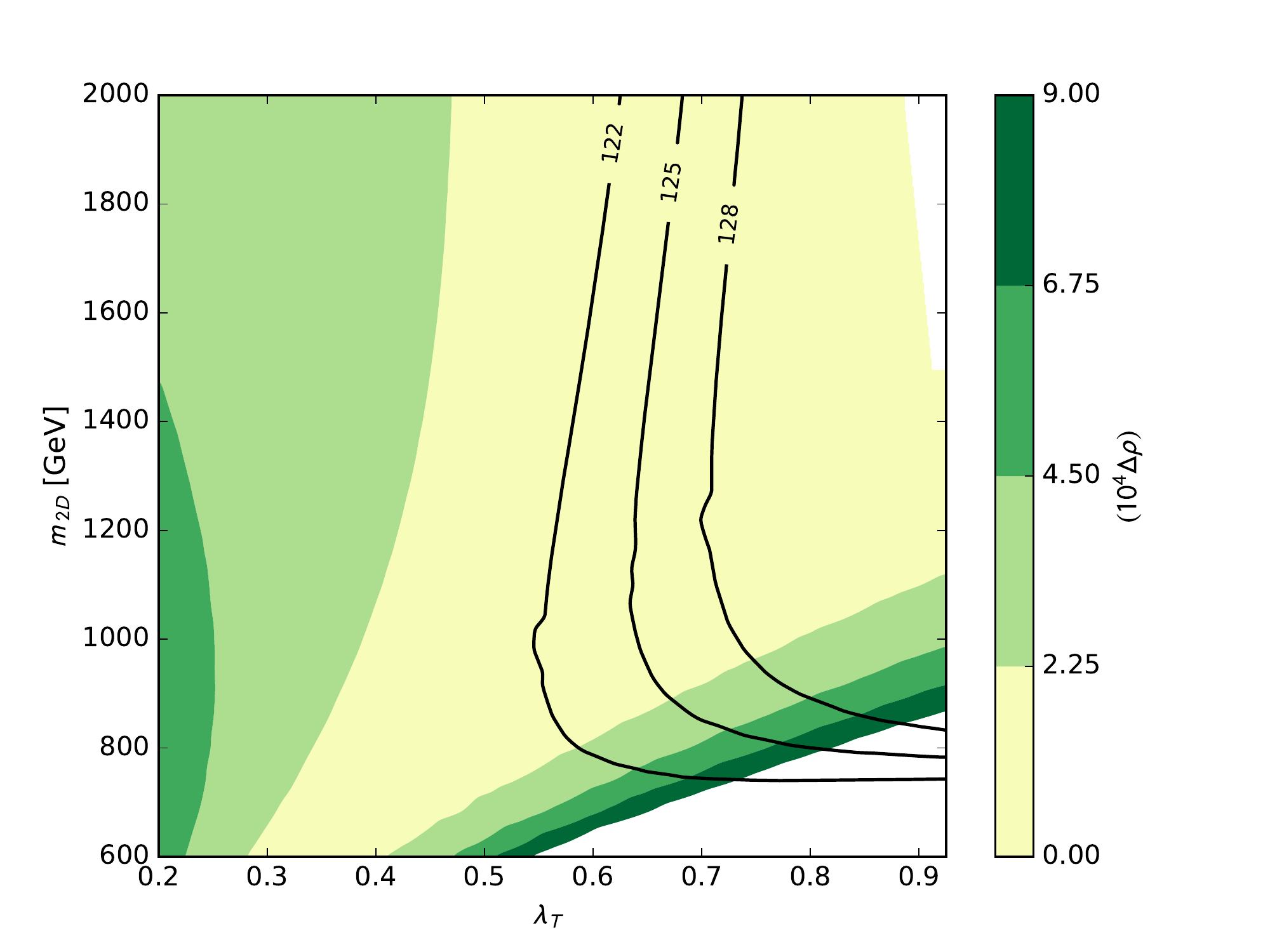}
\caption{ One-loop $(10^4 \cdot \Delta \rho)$ in scenario \Ra obtained from the benchmark point of Table~\eqref{benchmarkS1} by varying $\lambda_T$ and $m_{2D}$. We have taken $m_T = 450$ GeV. The black lines give the contours for $m_H = 122,125$ and $128$.}
\label{fig:deltarho}
\end{center}
\end{figure}

Finally, the VEV of $T$ gives a contribution to the $W$ boson mass and the electroweak precision data give bounds on it. One must examine the induced correction $\Delta \rho$ to the Veltman $\rho$-parameter:
\begin{align}
\rho \equiv \frac{M_W^2}{c_{\theta_W}^2 M_Z^2} = 1+ \Delta \rho \ ,
\end{align}
with $\Delta \rho$ given analytically at tree-level by (\! \cite{goodsell_two-loop_2013})
\begin{align}
\Delta \rho \sim \frac{4 v_T^2}{v^2} \ ,
\end{align}
where $v$ is the usual Standard Model Higgs VEV. In order to be below the experimental constraints, we need $\Delta \rho \lesssim (4.2 \pm 2.7) \times 10^{-4}$,  (~\cite{goodsell_two-loop_2013} -- see also~\cite{Bertuzzo:2014bwa,Beauchesne:2014pra} --). At tree level, we have
\begin{eqnarray}
v_T & \simeq &  \frac{v^2}{2\tilde{m}^2_{TR}} \ \ \left[ - g   
m_{2D}  c_{2\beta} -{\sqrt{2}} \tilde{\mu}   \lambda_T \right] \ ,
\end{eqnarray}
with $\tilde{m}^2_{TR}= m_T^2+4 m^2_{2D}+ B_T$, therefore, small $\Delta \rho$ require large triplet Dirac and soft masses. This requirement can often be at odd with naturalness which prefers smaller triplet  masses. Indeed, radiative corrections induced by the adjoint triplet scalars to $m^2_{H_{u,d}}$ are~\cite{goodsell_two-loop_2013}:
\begin{align}
\delta m^2_{H_{u,d}} \supset - \frac{1}{16 \pi^2} ( 2 \lambda_T^2 m_T^2) \textrm{log} \left\lbrace \frac{\Lambda}{\textrm{TeV}} \right \rbrace \ ,
\end{align}
with $\Lambda$ the UV cut-off, $m^2_{H_{u,d}},  m_T^2$ the squared masses for Higgses and scalar triplet $T$, and $\lambda_T$ the coupling defined in~\eqref{WDG}. For $\Lambda$ at the Planck scale, requiring a fine-tuning $\Delta_{T} = \delta m_H^2 / m_H^2$ better than $10\%$ finally gives us
\begin{align}
 m_{T} \lesssim \frac{1}{\lambda_T}  450 \textrm{ GeV} \ .
\end{align}
In Figure~\ref{fig:deltarho}, we show the allowed region for $\lambda_T$ and $m_{2D}$ for $m_T = 450$ GeV. $\Delta \rho$ has been obtained at one-loop using the Spheno~\cite{porod_spheno_2003,porod_spheno_2012} code generated by \SARAH~(see ref.~\cite{staub_sarah_2008,staub_automatic_2011,staub_superpotential_2010,staub_sarah_2013,staub_sarah_2014}). We see that the Higgs mass prefer large values of $\lambda_T$ but that the following three requirements are perfectly compatible: (1) a $125$ GeV Higgs, (2) a natural mass for the triplet and (3) a parameter $\Delta \rho$ smaller than the current constraints.

%% file: Octets.tex
\begin{figure}[!htbp]
\begin{center}
\includegraphics[width=0.65\textwidth]{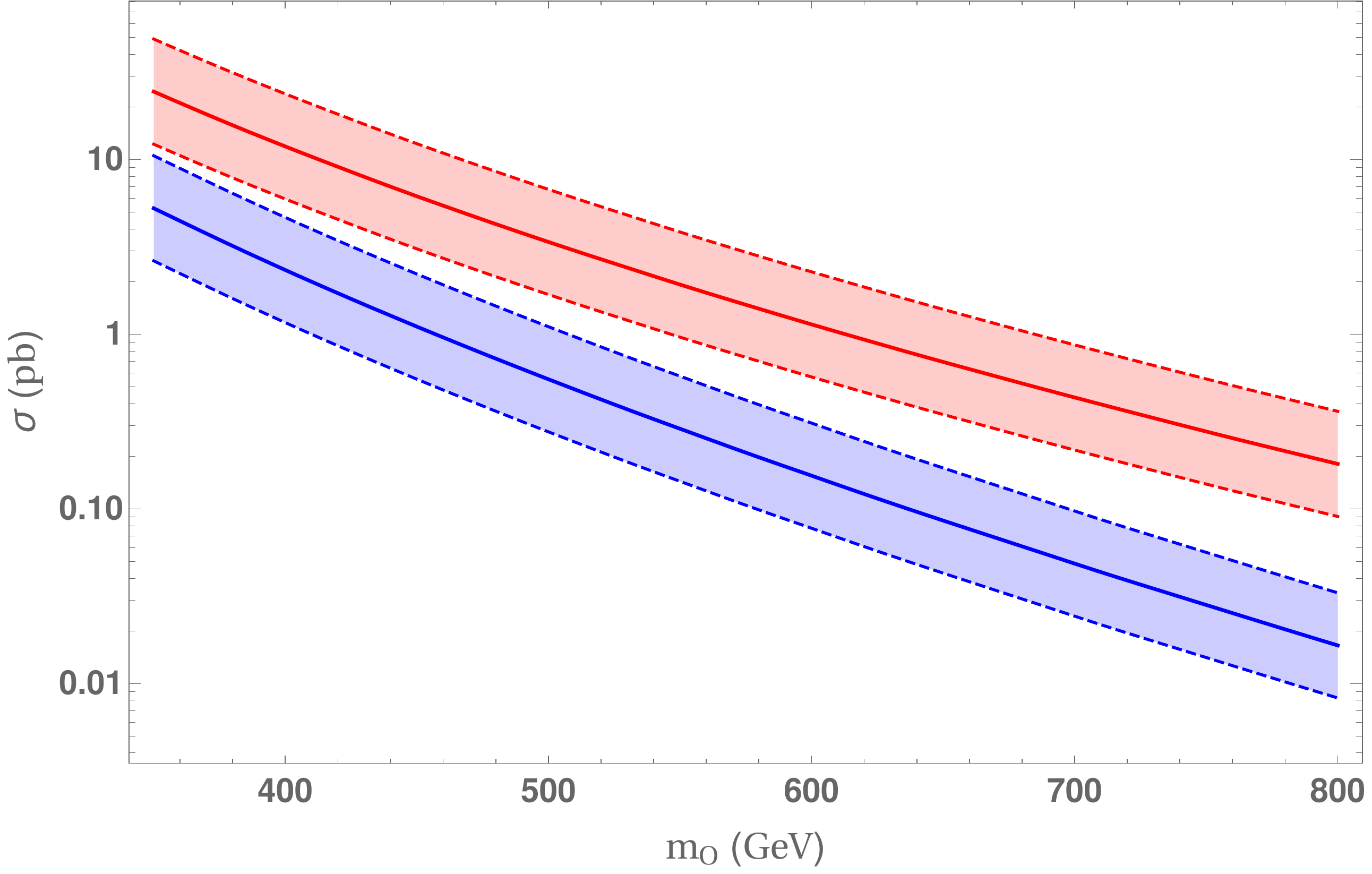}
\caption{Pair production cross-section of octets at tree-level, at 8 TeV (blue, lower curves) and 13 TeV (red, upper curves). The bands indicate a variation of a factor of $2$ each way relative to the values obtained in {\tt MadGraph}.  }
\label{fig:OctetX}
\end{center}
\end{figure}

In this work we shall be interested in the case when either the scalar or pseudoscalar colour octets are lighter than a TeV. Even though such light scalars should be copiously produced in pairs at both $8$ and $13$ TeV, as shown in figure \ref{fig:OctetX}, their decays are loop suppressed and this inhibits single production. 

Since current limits place all squarks above about $800$ GeV, then, as first discussed in \cite{Choi:2008ub,Plehn:2008ae}, the octets decay only to gluons and quarks -- in particular almost entirely top quarks. This means that the possible signatures are four jets, dijet/ditop searches, and four tops. Up until relatively recently the constraints on them were rather weak, with dijets providing no constraint, and a mild constraint from ditops \cite{Chatrchyan:2013lca}. However, now the four top channel is particularly important: \cite{Khachatryan:2014sca} placed a limit of $32$ fb at $8$ TeV, and \cite{ATLAS:FOURTOPS} found $370$ fb for Standard-Model-like kinematics, or $140$ fb with and EFT pointlike interaction, at $13$ TeV. 

To interpret the implications of these searches for our model, we could in principle do a full recasting along the lines of \cite{Beck:2015cga}; however, for simplicity we shall consider instead the cross-section times branching ratio approach, taking the most conservative values of twice the tree-level cross-section (i.e. a K-factor of $2$) and a limit at $13$ TeV of $140$ fb. To compute the branching ratio into four tops, we require the widths into gluons and tops; while expressions were given for these originally in \cite{Choi:2008ub,Plehn:2008ae}, those papers used \emph{complex} octets, which is not appropriate for our case where the necessarily large ($\gtrsim 2$ TeV) gluino mass causes a large splitting. Instead we require the expressions presented in \cite{Goodsell:2014dia}, which we shall not reproduce here but to which we refer the reader. 

The first important observation is that the pseudoscalar octet \emph{does not couple} to gluons, and so pair production of pseudoscalar octets yields only four-top events, and by our above criteria excludes pseudoscalars below about $880$ GeV \emph{by the $13$ TeV data.} These are therefore less interesting for our analysis.

\begin{figure}[!htbp]
\begin{center}
\includegraphics[width=0.49\textwidth]{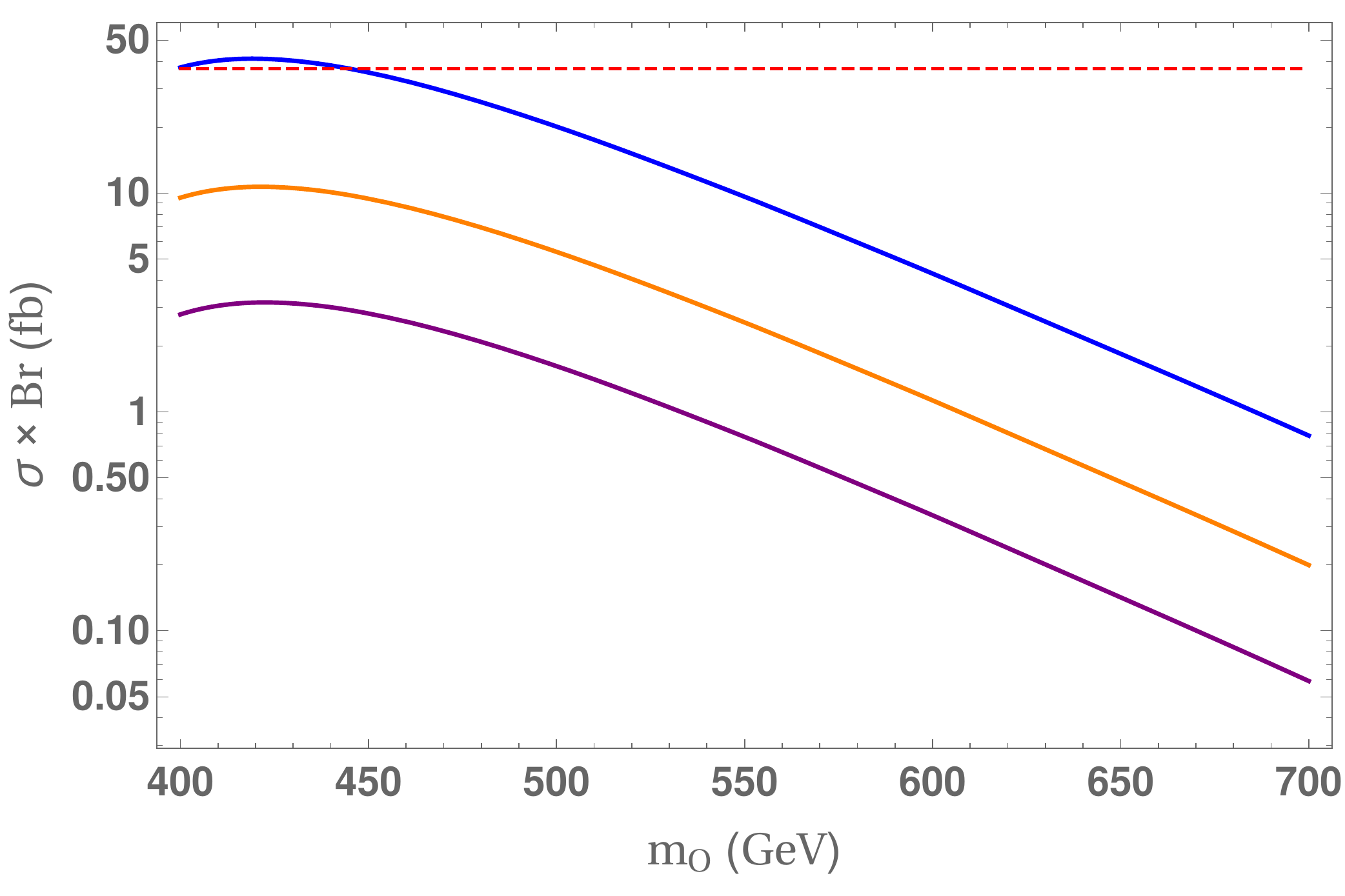}
\includegraphics[width=0.49\textwidth]{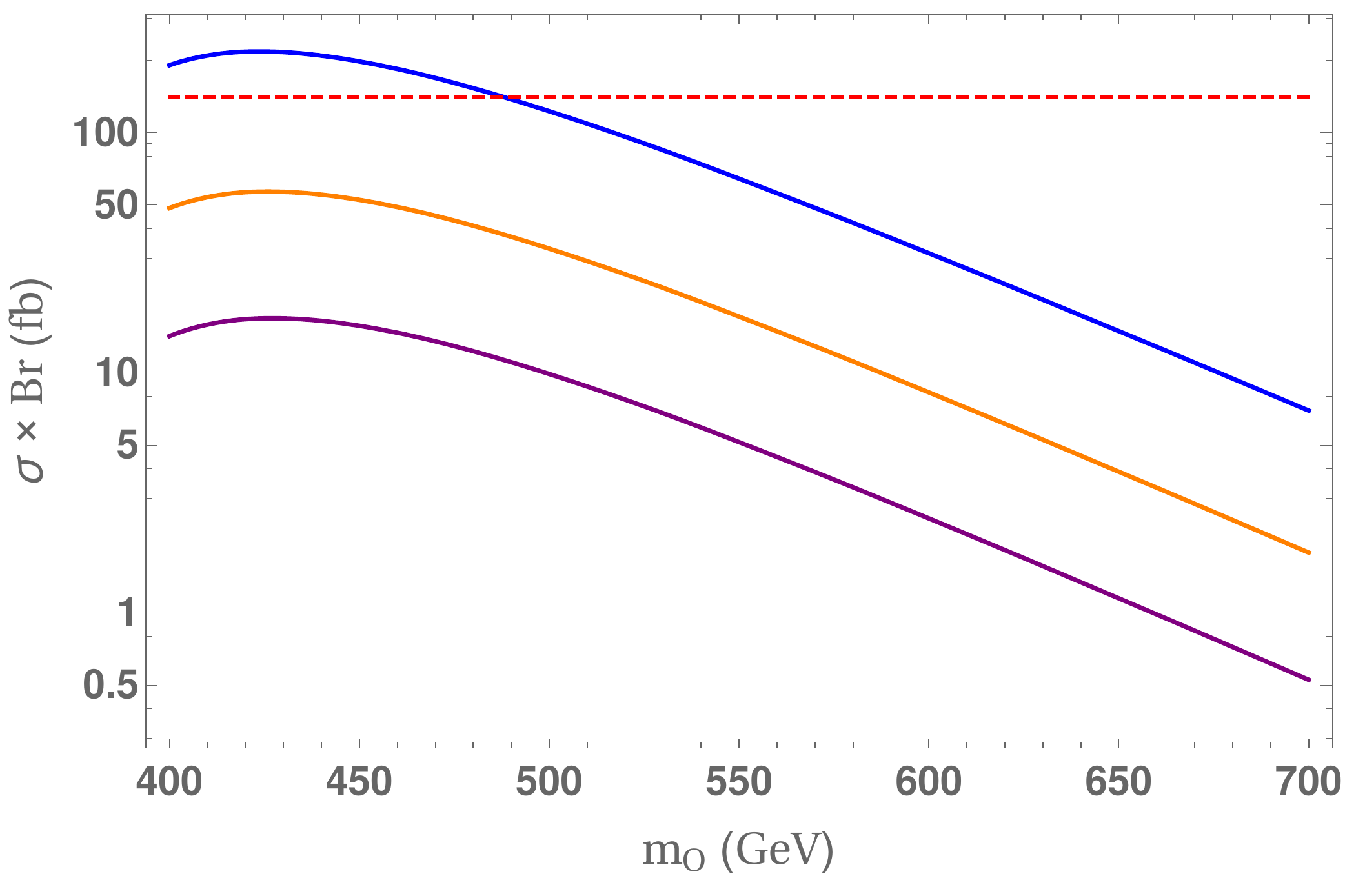}
\caption{Four-top production times branching ratio from scalar colour octets as a function of the octet mass, for gluino masses of $2.5$ TeV (upper curve, blue), $3$ TeV (middle curve, orange) and $3.5$ TeV (lower curve, purple). The experimental limit is shown as the dashed red horizontal line. The left plot is computed for $\sqrt{s} = 8$ TeV, and right is for $\sqrt{s} = $ 13 TeV.  }
\label{fig:Octettttt}
\end{center}
\end{figure}

On the other hand, the scalar octet couples to squarks via its D-term coupling, and so couples to gluons. Since it couples to \emph{all} coloured squarks, this can potentially be large. However, to be very conservative, we show production times branching ratio of four-tops via scalar octets in figure \ref{fig:Octettttt} at $8$ and $13$ TeV with the limits shown using a K-factor of $2$, as we vary the octet mass and for three different values of the Dirac gluino mass, where the first two generations of squarks are decoupled (i.e. heavy and degenerate). To produce these, we take left-handed stops and sbottoms of $1200$ GeV, right-handed stops of $800$ GeV, and decoupled right-handed sbottoms (at $4$ TeV). We neglect all squark mixing (which is a good approximation in this model). Since the couplings involve a cancellation between left- and right-handed squarks, this is very conservative: if we took heavier left-handed squarks, we would enhance the gluon rate relative to the top rate (because it has a contribution from sbottoms as well as stops) weakening the bounds. 

We conclude that for $2.5$ GeV gluinos, the octet scalars must be heavier than $500$ GeV; but for $3$ TeV gluinos there is no constraint. 

%% file: Unification.tex
The field content of the MDGSSM, and more precisely the two pairs of vector-like electrons $\hE$ and $\hEp$ as well as the doublet $R_u, R_d$, have been chosen to have one-loop unification by completing the $\mathbf{8}_0+\mathbf{3}_0+\mathbf{1}_0$ set of adjoint multiplets into a complete GUT representation of $(SU(3))^3$ (see~\cite{Benakli:2014cia}). We have furthermore checked numerically that gauge couplings remain safely perturbative at two-loops up to the GUT scale, consistently with the results of~\cite{Benakli:2014cia}.

% , albeit its precision decreases when increasing the squarks mass. This effect is driven by the one-loop conversion\luc{I am right on this ?} between the experimentally measured gauge couplings and their running values in the $\overline{DR}$ scheme at the SUSY scale. Since this scale is traditionally given by the geometrical mean of the squark masses, large squarks implies large corrections and we expect to recover a more precise unification if two-loop conversion formulas were used.

Once the GUT scale is determined, we require perturbation theory to be valid up to the GUT scale. We choose as perturbativity requirement that all Yukawa couplings should remain smaller than $\sqrt{4\pi}$. As we will see now, this gives strong constraints on the Yukawa couplings. At one-loop,  the beta functions for $\lambda_{SE}, \lambda_{SR}, \lambda_{SO}, \lambda_S$ and $\lambda_T$ form a coupled system given by:
\begin{align*}
 \beta_{\lambda_S} &=~  \frac{1}{16\pi^2} \lambda_S [ 4 \lambda_S^2 +  3 \lambda_T^2  + 2 \lambda_{SR}^2  + 2\lambda_{SE}^2 + 4 \lambda_{SO}^2 - \frac{3}{5} g_1^2  - 3 g_2^2 +3 y_t^2+ \dots] \\
 \beta_{\lambda_T} &=~  \frac{1}{16\pi^2} \lambda_T [ 2 \lambda_S^2 + 4 \lambda_T^2 - \frac{3}{5} g_1^2  - 7 g_2^2 +3 y_t^2 \dots] \\
 \beta_{\lambda_{SE}} &=~  \frac{1}{16\pi^2} \lambda_{SE} [ 2 \lambda_S^2 + 4 \lambda_{SE}^2 + 2 \lambda_{SR}^2 +  4 \lambda_{SO}^2 - \frac{12}{5} g_1^2 + \dots ] \\
 \beta_{\lambda_{SR}} &=~  \frac{1}{16\pi^2} \lambda_{SR} [ 2 \lambda_S^2 + 2 \lambda_{SE}^2 + 4 \lambda_{SR}^2  + 4 \lambda_{SO}^2  - \frac{3}{5} g_1^2  - 3 g_2^2+ \dots ] \\
 \beta_{\lambda_{SO}} &=~  \frac{1}{16\pi^2} \lambda_{SO} [ 2 \lambda_S^2 + 4 \lambda_{SE}^2 + 2 \lambda_{SR}^2 +  6 \lambda_{SO}^2 - 12 g_3^2 + \dots ] \ ,
\end{align*}
where the dots contain the contributions from the other couplings. Before studying this system numerically, we point out some peculiarities of these expressions:
\begin{itemize}
 \item The gauge couplings contribute negatively to the beta function, increasing the stability. In particular, $\lambda_{SO}$ is strongly stabilised.
\item In the limit $\lambda_S \rightarrow  0$, $\lambda_T$ completely decouples from the other Yukawa couplings.
\item The perturbativity of the coupling $\lambda_S$ will be critical as: (1) the gauge couplings and top Yukawa already give a positive contribution $\sim 1.1$ to its beta function; (2) all the other Yukawas feed intro its beta function and conversely $\lambda_S$ feeds into all the beta functions.
\end{itemize}
% Considering the couplings one-by-one non-zero and neglecting the others, one get the simple equations:
% \begin{align*}
%  \beta_{\lambda_S} &=~  \frac{1}{16\pi^2} 4 \lambda_S^3  \\
%  \beta_{\lambda_T} &=~  \frac{1}{16\pi^2} 4 \lambda_T^3  \\
%  \beta_{\lambda_{SE}} =  \beta_{\lambda_{SR}} &=~  \frac{1}{16\pi^2}  6 \lambda_{SE}^3  \\
% \end{align*}
% which are easily solved to obtain the usual Landau poles bounds for $\lambda$, one of the previous coupling,  
% \begin{align}
%  \lambda^2 < \frac{16 \pi^2}{2 \beta_\lambda}  \frac{1}{\ln (M_{GUT} / M_{SUSY})} \ ,
% \end{align}
% where $M_{GUT}$ is the GUT scale and $M_{SUSY}$ the SUSY from which we start the running. Depending on the SUSY scale and on the GUT scale, this leads to the rough criterium
% $\lambda_S,\lambda_T \lesssim 0.8$ and $\lambda_{SE}, \lambda_{SR}  \lesssim 0.65$ depending on the precise unification scale (notice again the normalisation factor on $\lambda_T$).

% In our case as we want to take $\lambda_{SE}$ as large as possible (and generically $\lambda_T$ large to increase the tree-level Higgs mass). Furthermore, it also neglects the effect of the gauge couplings which can be sizeable, especially for $\lambda_{SO}$.

We have numerically constrained the initial values for $\lambda_{SE}, \lambda_{SR}, \lambda_{SO}, \lambda_S$ and $\lambda_T$ at the low scale (SUSY scale), so that they remain perturbative up to the GUT scale. We use the two-loop RGEs generated by the public code \SARAH~(see ref.~\cite{staub_sarah_2008,staub_automatic_2011,staub_superpotential_2010,staub_sarah_2013,staub_sarah_2014} and ref.~\cite{goodsell_two-loop_2013}). 

\begin{figure}[!htbp]
\begin{center}
\includegraphics[width=0.49\textwidth]{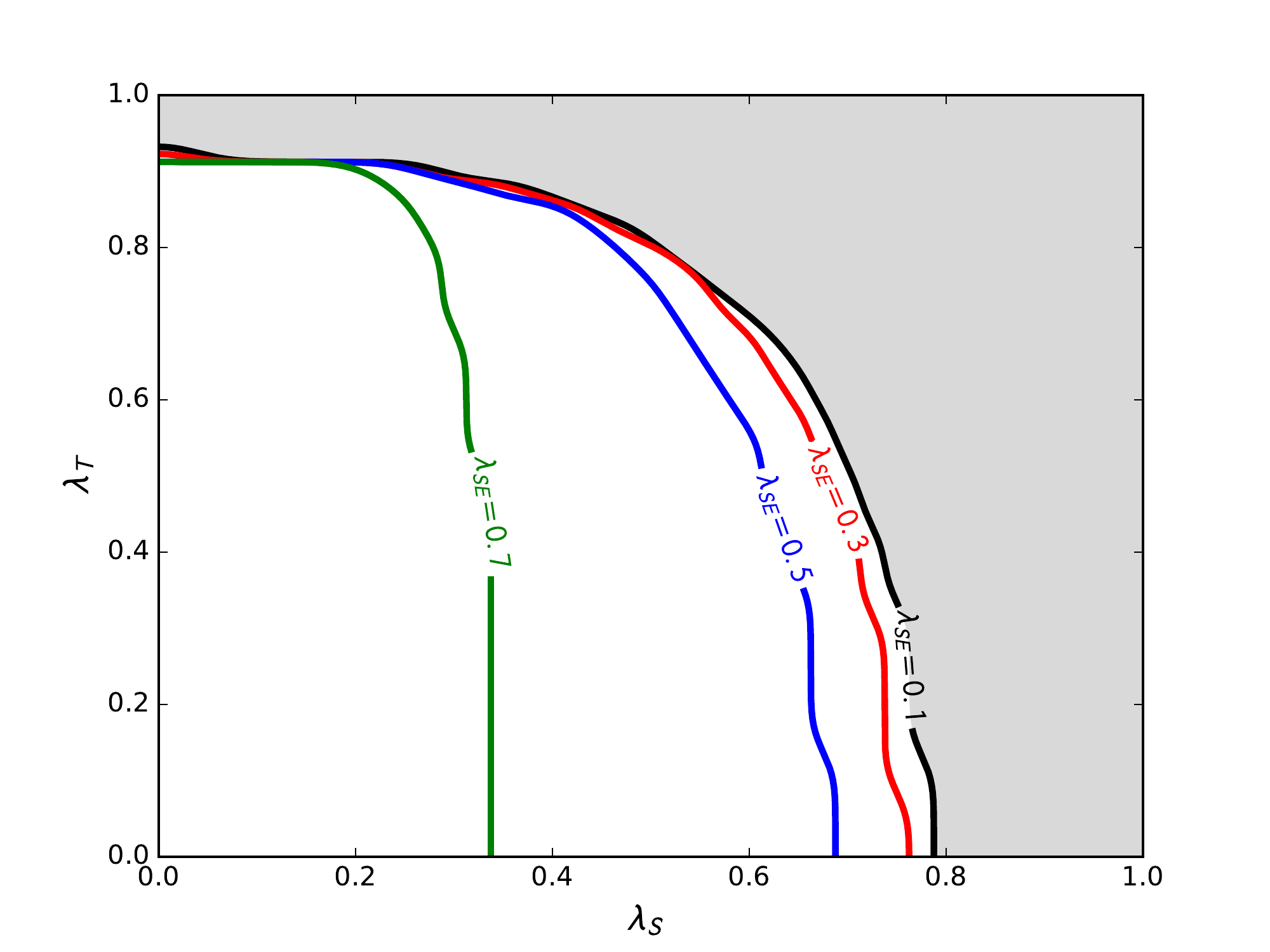}
\includegraphics[width=0.49\textwidth]{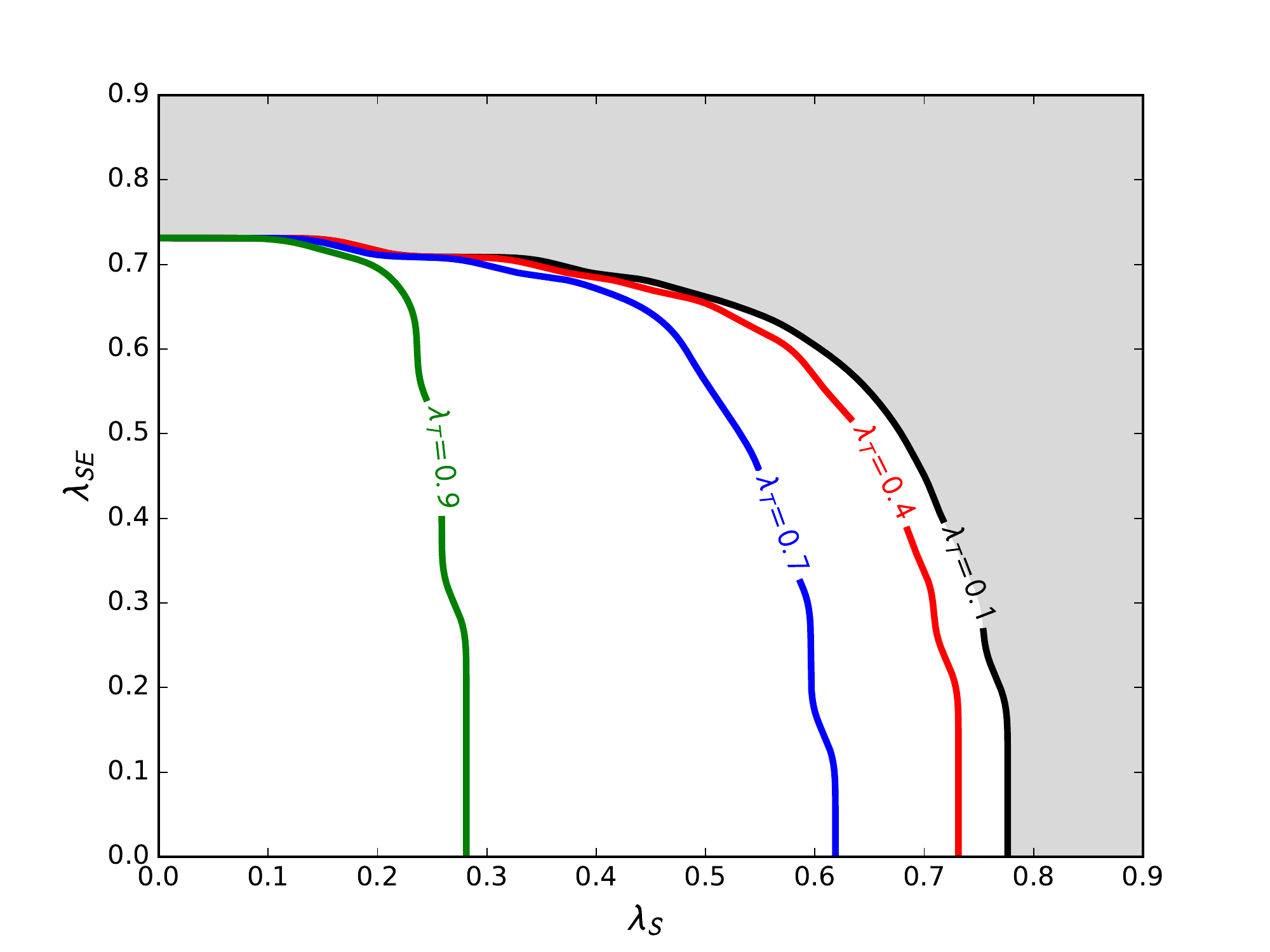}
\caption{Perturbativity bounds on our model, around the first benchmark point from Table~\ref{benchmarkS1}, obtained from the requirement that no couplings overtake $\sqrt{4\pi}$ before the GUT scale. We consider $\lambda_{SR} = \lambda_{SE}$. \textbf{Left plot:} Bounds for (from left to right) $\lambda_{SE}=0.7,0.5,0.3,0.1$ in the $\lambda_S / \lambda_T$ plane, all points above the curves are excluded.  \textbf{Right plot:} Bounds for (from left to right) $\lambda_{T}=0.9,0.7,0.4,0.1$ in the $\lambda_S / \lambda_{SE}$ plane, all points above the curves are excluded.  }
\label{fig:LandauR}
\end{center}
\end{figure}

\begin{figure}[!htbp]
\begin{center}
\includegraphics[width=0.49\textwidth]{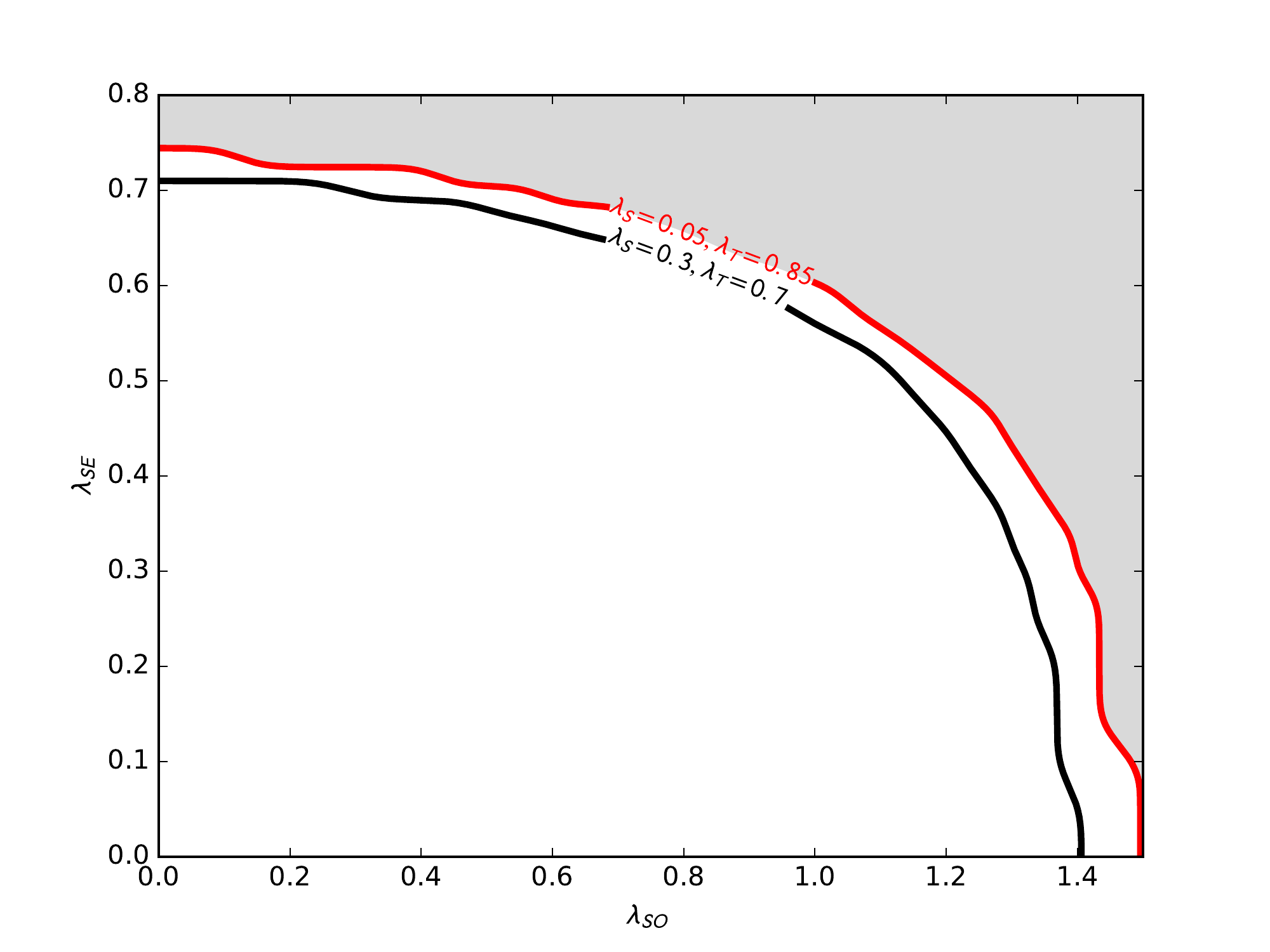}
\includegraphics[width=0.49\textwidth]{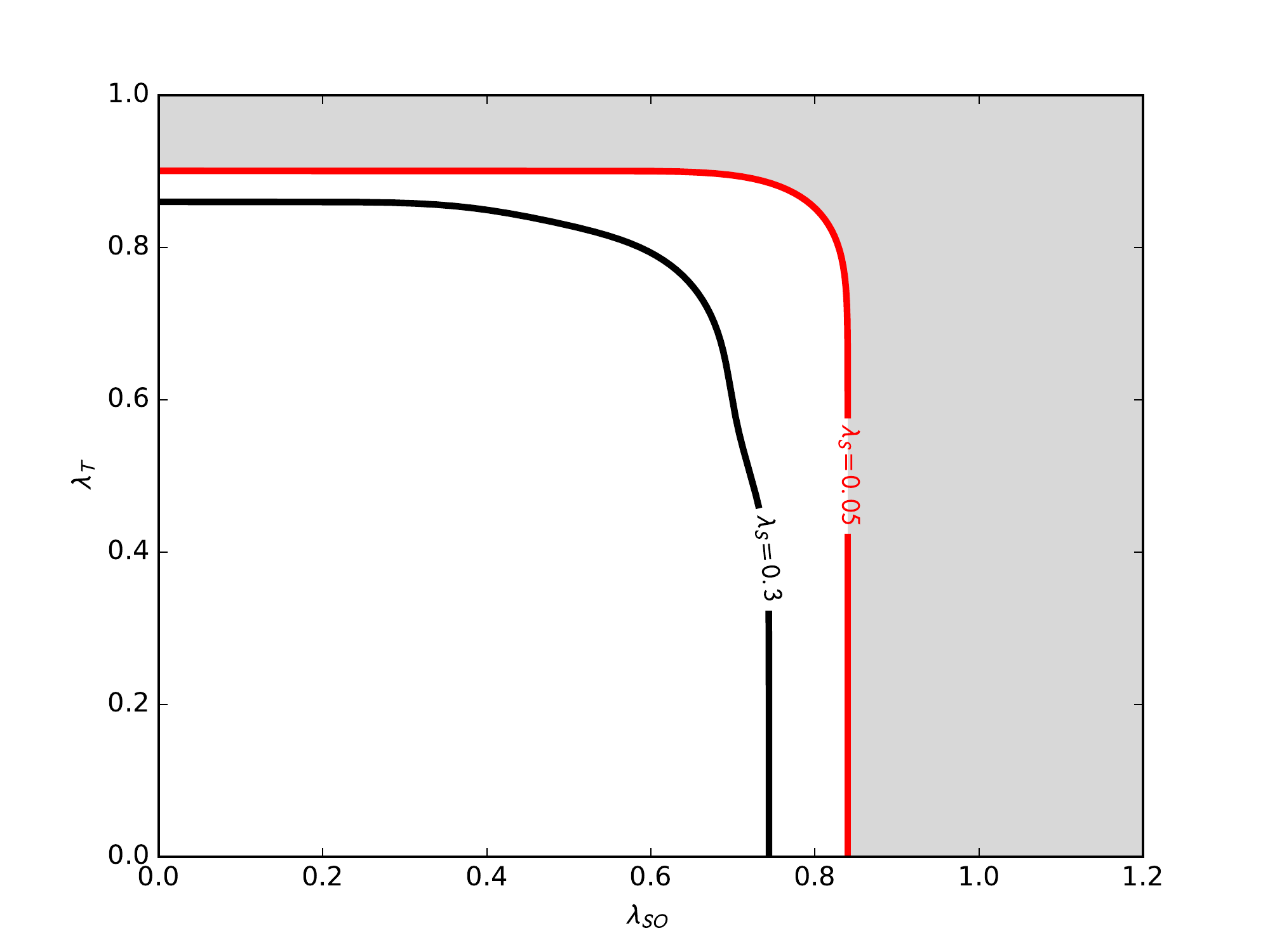}
\caption{Perturbativity bounds on our model around the first benchmark point from Table~\ref{benchmarkS1}, obtained from the requirement that no couplings overtake $\sqrt{4\pi}$ before the GUT scale. We consider $\lambda_{SR} = \lambda_{SE}$. \textbf{Left plot:} in the $\lambda_{SO} / \lambda_{SE}$ plane with from left to right $(\lambda_{S} = 0.3, \lambda_T = 0.7)$ and $(\lambda_{S} = 0.05, \lambda_T = 0.85)$; all points above the curves are excluded. \textbf{Right plot:} in the $\lambda_{SO} / \lambda_{T}$ plane with from left to right $(\lambda_{S} = 0.3, \lambda_{SE} = 0.65)$ and $(\lambda_{S} = 0.05, \lambda_{SE} = 0.65)$; all points above the curves are excluded.  }
\label{fig:LandauRV}
\end{center}
\end{figure}

In Figure~\ref{fig:LandauR}, we study the case of $\lambda_{SO} = 0$, which will be relevant for the two R-conserving scenarios \Ra and \Rb. The perturbativity bounds are shown in the planes $\lambda_S / \lambda_{SE}$ and $\lambda_S / \lambda_{T}$. As expected, we obtain the strongest constraints for $\lambda_S$, especially in the large $\lambda_{SE}$ case, which is the one of interest in this paper. Furthermore, we recover that for $\lambda_{S} \rightarrow 0$, $\lambda_T$ is insensitive to the other Yukawa couplings.
Adding the parameter $\lambda_{SO}$ further constrains the Yukawa couplings. This is shown in Figure~\ref{fig:LandauRV} where we present the perturbativity bounds on $\lambda_{SE}$ and $\lambda_{SO}$ for various values of $\lambda_{S}$ and $\lambda_{T}$. We see that for $\lambda_{SO} \sim 0.65$, one should take $\lambda_{SE} < 0.65$ to be safely perturbative. Furthermore, as expected from the one-loop beta functions, $\lambda_{SO}$ has an increased stability thanks to the strong gauge coupling contribution, allowing values up to $1.4$ for low $\lambda_{SE}$. Notice in the right-hand plot of Figure~\ref{fig:LandauRV} that in the limit $\lambda_S \rightarrow 0$, we recover that $\lambda_T$ decouples from the other Yukawas.

%% file: VacuumStudy.tex
We now turn to the constraints from vacuum stability; since we have significant trilinear scalar couplings then this is of crucial importance. The tree-level scalar potential can be decomposed into four main contributions:
\begin{align}
 V = V_{g} + V_{W} + V_{\rm soft} + V_{\rm hard}\ ,
\end{align}
with $V_{g}$, containing the D-term contributions, $V_W$ the superpotential contributions and $V_{\rm soft}$ the soft SUSY-breaking terms. The final term $V_{\rm hard}$ consists of ``hard'' dimensionless quartic terms that are generated at the SUSY-breaking scale and look like hard SUSY-breaking terms discussed in section \ref{SEC:TRILINEARS}. 

We have 
\begin{align*}
 V_{g} = \frac{1}{2} D_1^2 + \frac{1}{2}  D_{2a}D_2^a + \frac{1}{2} D_{3a} D_3^a
\end{align*}
where 
\begin{align*}
D_1 & =   - 2 m_{1D}  S_R +D_Y^{(0)} \qquad \textrm{with}  \qquad D_Y^{(0)}=  - g'\sum_{j} Y_j \varphi_j^\dagger \varphi_j \label{Electroweak:DtermsS}\\ 
D^a_2 & =   - \sqrt{2} m_{2D} (T^a+ T^{a\dagger}) +D_2^{a(0)} \qquad \textrm{with}  \qquad D_2^{a(0)}=  -g_2 \sum_{j}  \varphi_j^\dagger  M_j^a \varphi_j \\
D^a_3 & =   - \sqrt{2} m_{3D} (O^a+ O^{a\dagger}) +D_3^{a(0)} \qquad \textrm{with}  \qquad D_3^{a(0)}=  -g_3 \sum_{j}  \varphi_j^\dagger  M_j^a \varphi_j  \ .
\end{align*}
where $\varphi_j $ are the scalar components of the matter chiral superfields, possibly in the adjoint representation and $M^a_j$ is the matrix of the gauge representation of $\varphi_j $. Let us leave aside the triplet contribution (we are considering a heavy triplet and therefore expect a near-zero VEV for it) and focus on the singlet and octet terms. Similarly, we will leave aside the squarks contribution as we are not considering large $A$ terms and therefore do not expect them to acquire a color-breaking VEV. We have then
\begin{align*}
D_1^{(0)} &= - \frac{g'}{2} (R_u^\dagger R_u - R_d^\dagger R_d) - g'(|\hE_{}i|^2 -|\hEp{}_i|^2 ) \\
D_2^{a(0)} &= - g_2 (R_u^\dagger \frac{\sigma^a}{2} R_u + R_d^\dagger \frac{\sigma^a}{2}R_d) \\
D_3^{a(0)} &= - g_3 O_b^\dagger (T^a)^{bc} O_c \ ,
\end{align*}
with $(T^a)^{bc} = (-i f^{abc})$ and $f^{abc}$ the $SU(3)$ structure constants.

We now turn to the superpotential contributions (we suppress the $i$ indices for $\hEp{}_i$ and $\hEp{}_j$ and the ``$\cdot$'' denotes $SU(2)$ indices contraction by $\epsilon$ tensors) and find:
\begin{align*}
 V_{W}  ~=~ & \mu_r^2 (R_u^\dagger R_u + R_d^\dagger R_d) + \mu_E^2 (|\hE|^2 + |\hEp|^2) \\
& + \lambda_{SE}^2 \left[ |\hEp \hE|^2 + |S|^2 (|\hE|^2 + |\hEp|^2) \right] + \lambda_{SR}^2 \left[ |R_u \cdot R_d|^2 + |S|^2 (|R_u|^2 + |R_d|^2 ) \right]   \\
& 
\end{align*}

The only ``hard'' SUSY-breaking terms that will be of relevance to us will be a quartic octet coupling:
\begin{align}
V_{\rm hard} \equiv& \frac{\lambda_O}{4} |O^a|^4 + \lambda_{SO}^H |S|^2 |O^a|^2 
\end{align}
which is of course not the only such possible term but is the most important.

After adding the soft and hard SUSY-breaking terms, we obtain 
\begin{align}
 V = V_E + V_{SE} + V_{SR} + V_{S} + V_R + V_O + V_{SO} \ ,
\end{align}
with 
\begin{align}
 V_E ~=~ &  (m_E^2+\mu_E^2) (|\hE|^2 + |\hEp|^2) + \lambda_{SE}^2 |\hEp \hE|^2 + \frac{g'^2}{2}(|\hE|^2 -|\hEp|^2 )^2 + B_E ( \hE \hEp + h.c.) \nn\\[0.6em]
 V_S ~\supset~&  m_S^2 |S|^2 +2 m_{1D}^2 S_R^2 + \frac{1}{2} B_S ( S^2 + h.c.) \nn\\[0.6em]
 V_R ~\supset~&  (m_R^2+ \mu_r^2) (R_u^\dagger R_u + R_d^\dagger R_d) + \lambda_{SR}^2  |R_u \cdot R_d|^2+ B_R ( R_u \cdot R_d + h.c.)\nn \\
      &  +\frac{1}{8} \left[  \gp^2 \ ( R_u^\dagger R_u - R_d^\dagger R_d )^2  + g_2^2  (R_u^\dagger \frac{\sigma^a}{2}R_u + R_d^\dagger \frac{\sigma^a}{2}R_d)^2 \right] \nn\\[0.6em]
 V_O~\supset~ &  2 m_O^2 \textrm{tr}(O^\dagger O) + 2 m_{3D}^2 \textrm{tr}(O_R^\dagger O_R) + (B_O \textrm{tr}(OO)+ h.c.) \nn \\
       &  + \frac{g_3^2}{2} \left[ (O_b^\dagger (T^a)^{bc} O_c) (O_b^\dagger (T_a)^{bc} O_c) \right] + \sqrt{2} g_3 m_{3D} (O + O^\dagger)^a O_b^\dagger (T_a)^{bc} O_c \nn \\
& + ( T_O \mathrm{tr} (O^3) + h.c. )+ \frac{\lambda_O}{4} |O^a|^4 \ ,
\end{align}
and the mixed contributions 
\begin{align*}
 V_{SE} & ~=~  2\gp m_{1D} S_R (|\hE|^2 -|\hEp|^2)  +\lambda_{SE}^2 |S|^2 (|\hE|^2 + |\hEp|^2 ) + T_{SE} (S \hE \hEp + h.c.) \\[0.6em]
 V_{SR} & ~=~  \gp m_{1D} S_R (R_u^\dagger R_u - R_d^\dagger R_d)  + \lambda_{SR}^2  |S|^2 (R_u^\dagger R_u + R_d^\dagger R_d)^2 + T_{SR} (S R_u \cdot R_d + h.c.) \\[0.6em]
 V_{SO} & ~=~ T_{SO} ( S\textrm{tr}(OO)+ h.c.) + \lambda_{SO}^H |S|^2 |O^a|^2\ .
\end{align*}

\subsubsection{Charge-breaking minima}

First, we investigate the $S, \hE, \hEp$ sector, which can drive charge-breaking minima when they all acquire a vev. The relevant tadpoles (for just one pair of $\hE, \hEp$) are
\begin{align}
 \frac{\partial V}{\partial \ov{S}} ~\simeq~ & m_S^2 S + \ov{B}_S \ov{S} + m_{DY}^2 (S + \ov{S}) +   \gp \sqrt{2} m_{1D} (|\hE|^2 -|\hEp|^2) + \lambda_{SE}^2 S (|\hE|^2 + |\hEp|^2) + T_{SE} \hE \hEp \nn \\[0.6em]
 \frac{\partial V}{\partial \ov{\hE}} ~=~ &  \hE ( m_E^2 + \mu_E^2 + \lambda_{SE}^2 |S|^2 + \gp  m_{1D} S_R )  + \hEp(B_E + T_{SE} S) + \lambda_{SE}^2 \hE |\hEp|^2 	\nn \\[-0.6em]
 &  +  \gp^2 \hE (|\hE|^2 - |\hEp|^2) + m_{\hE}^2 |\hE|^2 \nn \\
 \frac{\partial V}{\partial \ov{\hEp}} ~=~&  \hEp ( m_E^2 + \mu_E^2 + \lambda_{SE}^2 |S|^2 - \gp  m_{1D} S_R )  + \hE(B_E + T_{SE} S) + \lambda_{SE}^2 \hEp |\hE|^2 	\nn \\[-0.6em]
 & -  \gp^2 \hEp (|\hE|^2 - |\hEp|^2) \ . 
\end{align}
We have two limits relevant for this model: relevant for this paper is the case that $m_{DY}$ is not large, in which case the most dangerous direction is the ``classic'' D-flat direction $|\hE|^2 = |\hEp|^2$. When $S, \hE, \hEp$ develop vevs, we can decompose the complex fields into real and imaginary parts; without loss of generality we can put $\hE = \hEp \equiv \frac{1}{\sqrt{2}} E_R, S = \ov{S} \equiv \frac{1}{\sqrt{2}} s_R $. Solving then the equation for the singlet tadpole, we find the potential
\begin{align}
V\big|_{s_R} =& \frac{E_R^2}{4( \lambda_{SE}^2 E_R^2 + m_{SR}^2)} \bigg[ \lambda_{SE}^4 \big( E_R^2 + \frac{1}{2} ( 2 m_{ER}^2  + m_{SR}^2 - \hat{T}_{SE}^2 \big)^2 \nn\\
& \qquad -\frac{1}{4}\bigg(  \hat{T}_{SE}^4 - 2 \hat{T}_{SE}^2 ( m_{SR}^2 + 2 m_{ER}^2) + ( m_{SR}^2 - 2 m_{ER}^2)^2 \bigg) \bigg]
\end{align}
where we defined
\begin{align}
\hat{T}_{SE} \equiv& T_{SE}/\lambda_{SE} \nn\\
m_{SR}^2 \equiv& m_S^2 + B_S + 4 m_{DY}^2 \nn\\
m_{ER}^2 \equiv& m_{\hE}^2 + m_{\hEp}^2 + 2 B_{E} + 2 \mu_{E}^2 .
\end{align}
Clearly we observe that we have appearance of a charge-breaking vacuum if
\begin{align}
 \frac{T_{SE}^2}{\lambda_{SE}^2 } >  2 m_{ER}^2 + m_{SR}^2 \ . 
\end{align}
However, it is only lower than our vacuum if the weaker condition
\begin{align}
\frac{T_{SE}^2}{\lambda_{SE}^2}  >  2 m_{ER}^2 + m_{SR}^2 + 2 \sqrt{2} m_{ER} m_{SR}
\end{align}
is satisfied, or equivalently
\begin{align}
m_{ER} < \frac{1}{\sqrt{2}} \big( \frac{T_{SE}}{\lambda_{SE}} - m_{SR} \big). 
\end{align}
 The analagous constraints also apply for the pseudoscalar direction, and also for the $S, R_u, R_d$ sector.

\subsubsection{Colour-breaking minima}

A crucial part of our analysis is the presence of trilinear couplings of the singlet to the octet, which generate a coupling to gluons. However, just as the couplings to the selectron-like states allow charge-breaking minima, the octet scalar couplings permit colour-breaking minima. The analysis is identical for $O_R$ or $O_I$ with the opposite sign for $T_{SO}$, so let us choose $O_R$. The tadpole equations read
\begin{align}
0=&(m_S^2  + \frac{1}{2} \lambda_{SO}^H O_R^2 + \frac{1}{2} \lambda_{SO}^2 O_R^2) s_R - \frac{ O_R^2 T_{SO}}{2\sqrt{2}} \nn\\
0=&  O_R\big( m_{OR}^2 + \frac{\lambda_O + \lambda_{SO}^2}{4} O_R^2 + \frac{1}{2}(\lambda_{SO}^H +  \lambda_{SO}^2) s_R^2 -\frac{1}{\sqrt{2}} T_{SO} s_R\big)
\end{align}
where now $m_{OR}^2 \equiv m_O^2 + B_O + 4 |m_{D3}|^2$. We therefore see that the supersymmetric terms are equivalent to putting $\lambda_O = \lambda_{SO}^2, \lambda_{SO}^H = \lambda_{SO}^2$; in an analysis identical to the previous subsection we find that an additional vacuum exists when
\begin{align}
T_{SO}^2 >  (\lambda_O + \lambda_{SO}^2) m_{SR}^2 + 4 (\lambda_{SO}^H + \lambda_{SO}^2 )m_{OR}^2,
\end{align}
but that the minimum is only lower than the colour-preserving one when
\begin{align}
T_{SO}^2 > \big( 2 \sqrt{\lambda_{SO}^H + \lambda_{SO}^2} m_{OR} + \sqrt{\lambda_O + \lambda_{SO}^2} m_{SR} \big)^2,
\end{align}
or equivalently, when $\lambda_{SO} \ne 0$,
\begin{align}
m_{OR} < \frac{1}{2\sqrt{\lambda_{SO}^H + \lambda_{SO}^2}}  \big(  T_{SO} - m_{SR}  \sqrt{ \lambda_O + \lambda_{SO}^2} \big).
\end{align}

If we choose to break R-symmetry only in the Higgs sector via a $B_\mu$-term, then $\lambda_{SO} = 0$. In this case, we need to rely on $\lambda_{SO}^H$ and $\lambda_{O}$ only to stabilise the potential, leading to very strong constraints on the trilinear $T_{SO}$. For instance, if we have  $\lambda_{SO}^H , \lambda_{O} \sim \scr{O}(0.04)$ and a $400$ GeV scalar octet, the trilinear $T_{SO}$ must be smaller than $310$ GeV to ensure that the colour-preserving vacuum is stable.

%One should notice that the customary unstable direction $\hE = \hEp$ (where the potential is not protected by the quartic terms proportional to $\gp^2$), cannot be considered here. This is due to the presence of the Dirac mass term $\pm g^\prime  \hE m_{1D} S_R$ which prevent any solutions of the previous system in this direction. However, in the limit of interest for us $m_{1D} \ll T_{SE}$, we can in fact aproximate this term and take $\hE \simeq \hEp$ so that we can also neglect the $\gp^2$ quartic.

% In this limit, we find the solution to the previous system:
% \begin{align}
%  v_E^2 &= \frac{T_{SE}^2}{4 \lambda_{SE}^2} - (m_E^2 + \mu_E^2 + B_E^2) \\
% v_S &= -\frac{T_{SE}}{2 \lambda_{SE}} \nn \ .
% \end{align}
% so that we have appearance of a charge-breaking vacuum if
% \begin{align}
%  \frac{T_{SE}}{2 \lambda_{SE}}  > \sqrt{ m_E^2 + \mu_E^2 + B_E^2} \ . 
% \end{align}

% Similarly the condition to avoid appearance of charge-breaking vacuum in the $S, R_u, R_d$ sector is 
% \begin{align}
%  \frac{T_{SR}}{2 \lambda_{SR}}  > \sqrt{ m_R^2 + \mu_R^2 + B_R^2}  \ .
% \end{align}

%\subsection{Loop corrections}

%% file: Prelude.tex
\label{SUBSEC:Numerical}

While the MDGSSM has a large set of free parameters, the most relevant ones can be divided into three roughly independent sets controlling different features:
\begin{enumerate}
\item \textbf{Higgs and singlet masses and mixing}: $m_{1D}, m_S, B_S, \tan \beta, \mu, \lambda_S $ and $\lambda_T$. 
\item \textbf{Singlet decay/production amplitude to} $g g$: $T_{SO}, m_O, B_O$, $m_{\tilde{q}}$, where $m_{\tilde{q}}$ is the soft masses for right (or left)-handed squarks.
\item \textbf{Singlet decay amplitude to} $\gamma \gamma$: $T_{SE}, T_{SR}, \lambda_{SR}, \lambda_{SE}$ supplemented with soft masses and $B$ terms for the fields $\hE, \hEp, R_u$ and $R_d$.
\end{enumerate}
The first set is dedicated to reproducing the measured Higgs boson mass as well as a $750$ GeV scalar singlet. The value of $\lambda_S$ need to be adjusted to have a small mixing between both scalars, which is necessary both for the diphoton cross-section and for having $m_H \in [122,128]$.
The second set can is then used to enhance the production rate of singlet through gluon fusion. The trilinears $T_{SO}$ are crucial in this respect as they allow the scalar octet to participate in the loop-induced coupling $S g g$, greatly increasing the singlet production rate.
Finally, the last set of parameters is used to increase the diphoton amplitude. The superpotential Yukawa couplings $\lambda_{SE}$ and $\lambda_{SR}$ from~\eqref{WDG} are constrained to be below $0.7$ to avoid the appearance of Landau poles before the GUT scale. The trilinears are mainly constrained by enforcing that the scalar fields  $\hE, \hEp, R_u$ and $R_d$ does not get a charge-breaking vacuum expectation value.

We will investigate various scenarios that we can classify according to the presence or not of the R-violating terms~\eqref{WDG_RPV}:
\begin{itemize}
 \item R-symmetry preserving models (modulo, as discussed in section \ref{SEC:MDGSSM}, a $B_\mu$-term), which do not include the terms~\eqref{WDG_RPV} and have R-charges for the fake leptons such that only the superpotential couplings $\lambda_{E}$ and $\lambda_{SR}$ to $S$ are allowed. We distinguish the models
\begin{itemize}
\item \textbf{\Ra:} where we will consider large Dirac mass $m_{1D}$, so that the coupling to gluons proceeds through squarks loops.
\item \textbf{\Rb:} where instead consider small Dirac masses but light scalar octet, so that the coupling to gluons proceeds through scalar octet loops.
\end{itemize}
\item R-symmetry violating models, for which we can have additionally the terms~\eqref{WDG_RPV} and the trilinears $T_{SE}$ and $T_{SR}$. We consider
\begin{itemize}
 \item \Rva: A generalisation of scenario \Ra with $\lambda_{SO}$ and the trilinears $T_{SE}$ and $T_{SR}$ included.
 \item \Rvb: Similar to the model \Rva, but we further tolerate the presence of a Majorana gauginos mass terms. This allows to simultaneously produce the scalar $S_R$ and pseudo-scalar $S_I$ singlet and have a ``double-peaks'' resonance set-up.
\end{itemize}
\end{itemize}

In the following we shall present results of a numerical investigation of the parameter space of the MDGSSM for various scenarios. To do this we used the package \SARAH to produce \SPheno code to calculate the spectrum, production rate and decays. We created a new model file for the MDGSSM including the  adjoint couplings $\lambda_{SO}, T_{SO}$. However, we found that modifications to the \SPheno code were necessary:
\begin{enumerate}
\item We use pole masses instead of $\ov{\rm DR}$-masses for the selectrons and octet scalars in the calculation of loop couplings with the neutral scalars. This is because these masses are the most important for the gluon and photon couplings of our $750$ GeV candidate, and can differ by more than a factor of two; as described in \cite{Staub:2016dxq}, using the $\ov{\rm DR} $ masses is less accurate and so we employ pole masses just for these particles.
\item To facilitate our search for valid parameter points, we produced two different versions of the code. The first solves the tadpole equations for mass-squared parameters $m_{H_u}^2, m_{H_d}^2, m_S^2, m_T^2$ taking $v_S, v_T$ as inputs; while this is the appropriate choice for implementing the loop corrections to the scalar masses correctly, it is, however, difficult to choose the vacuum expectation values $v_S, v_T$ (since loop corrections can rapidly change the values of $m_S^2, m_T^2$ by several orders of magnitude). We therefore use this version of the code to check the results of our second code, which was specially modified to first solve the two-loop tadpole equations numerically for $v_S, v_T$, and then compute the tadpoles and masses using these values as inputs, solving for $m_S^2, m_T^2$ again along the same lines as the first code. While this is computationally expensive (computing the two-loop corrections to the neutral scalar tadpoles twice for each point) it is the most efficient way to correctly identify points -- and not miss points where, for example, $m_S^2$ may be identified as tachyonic at ``tree level.''  
\end{enumerate}

%% file: Pheno_Rsym.tex
Consider first the scenarios \Ra and \Rb where we include only the R-symmetry conserving adjoint couplings. Under these constraints, the singlet production proceeds mainly by gluons fusion through loops of squarks (controlled by  $\gp m_{1D}$) and (pseudo-)scalar octets (controlled by the trilinear $T_{SO}$). 

\subsubsection*{Squark-induced gluon fusion}

We start with scenario  \Ra and present in Table~\ref{benchmarkS1} a benchmark point satisfying all the previously-mentioned constraints while retaining a sizeable $\gamma \gamma$ cross-section.
\begin{table}[!ht]
\begin{center}
\begin{tabular}{c|c|c|c}
\hline
\small{}  &       Parameter          &  \Ra & \Rb              \\
\hline
\rule{0pt}{2.5ex} \small{\textbf{Higgs mass}}  & $\mu$   & $925$ GeV & $450$ GeV \\
 & $\mathrm{tan} \beta$   & $3$   & $5$  \\
 & $\lambda_{T}$   & $0.7$   & $0.85$  \\
 & $m_{T}$   & $500$ GeV  & $1000$ GeV  \\
\hline
\rule{0pt}{2.5ex} \small{\textbf{Singlet masses}}  & $m_{1D}$   & $1250$ GeV & $100$ GeV  \\
\small{\textbf{ and mixing}}& $m_{S}$   & $500$ GeV  & $775$ GeV \\
 & $B_{S}$   & $-2.44^2$ TeV$^2$ & $-200^2$ GeV$^2$  \\
 & $\lambda_{S}$   & $0.29$ & $0.05$    \\
\hline
\rule{0pt}{2.5ex} \small{\textbf{Singlet decay}}  & $T_{SO}$ & $200$ GeV &  $300$ GeV  \\
 \small{\textbf{/production amplitude} }& $m_{O}$   & $1300$ GeV & $1025$ GeV \\
 \small{\textbf{to} $g g$} & $ m_{\tilde{t}_R}$   & $500$ GeV  & $1200$ GeV\\
\hline
\rule{0pt}{2.5ex} \small{\textbf{Singlet decay amplitude} }  & $\lambda_{SR}=\lambda_{SE}$   & $0.7$ & $0.7$   \\
\small{\textbf{to} $\gamma \gamma$ }  & $m_E^2 =  m_{R_{u,d}}^2  $   & $10^2$ GeV$^2$ & $150^2$ GeV$^2$  \\
 & $ \mu_E =  \mu_{R_{u,d}}/1.4 $   & $325$ GeV  & $65$ GeV  \\
 & $ m_{\tilde{l} R} $   & $250$ GeV & $500$ GeV  \\
\hline
\hline
\rule{0pt}{2.5ex} \small{ \textbf{Outputs}  }  & Pole mass Higgs   &  $125.5$ GeV &  $124.9$ GeV  \\
 & Pole mass $S_R$  & $750.1$ GeV & $755.7$ GeV \\
&  Pole mass $O_I$/$O_R$ (\Ra/\Rb)  & $945.5$ GeV & $390.0$ GeV \\
&  Pole mass $\tilde{t}_R$  & $820.3$ GeV & $1165.0$ GeV  \\
&  Pole mass $\tilde{l}_R$  & $418$ GeV & $513$ GeV  \\
&  Pole mass $\tilde{\hE}$  & $397$ GeV & $382$ GeV  \\
 & \small{ $ \sigma ( \Stogg ) $ }    & $\mathbf{3.20}$ \textbf{fb}  & $\mathbf{3.18}$ \textbf{fb}  \\
 & \small{ $ \Delta \rho $ }    & $0.97\times 10^{-4}$ & $3.17\times 10^{-4}$    \\
 & \small{ $ v_S $ }    & $151.4$ GeV   & $643.5$ GeV  \\
\hline
\end{tabular}
\caption{Benchmark point for our scenario. We further have, $B_\mu = - 2.5^2$ TeV$^2$, the heavy left-handed squarks (as well as right-handed sbottom) have masses around $2.25$ TeV. The two first generation of right-handed squarks have masses at $975$ GeV (\Ra) or $1300$ GeV (\Rb), left-handed sleptons have masses at $1.5$ TeV. We have $m_{2D} = 1200$ GeV (\Ra) or $900$ GeV (\Rb), and $m_{3D} = 2.5 $ TeV (\Ra) or $3$ TeV (\Rb)}
\label{benchmarkS1} 
\end{center}
\end{table}

The main aspects of this scenario are the following:
\begin{itemize}
 \item We limit the R-symmetry breaking to the Higgs sector, and therefore choose R-charge of the fake fields to allow $\lambda_{SE},\lambda_{SR}$ superpotential terms but not trilinears $T_{SE}$/$T_{SR}$ and the corresponding B-terms.
 \item The loop coupling to gluons will proceed through squark loops, with singlet/squarks coupling enhanced by a large Dirac mass $m_{1D}$
 \item The loop coupling to photons have numerous contributions through loops of fake fields (both fermions and scalars) and sleptons. 
 \item Finally, because of the large Dirac mass, one need a sizeable negative $B_S$ to ensure that the scalar singlet has a mass of $750$ GeV.
 \end{itemize}
Notice that a satisfying feature of this scenario is that we do not need to fine-tune  the mass of the fields participating in the loop coupling beween $S$ and $\gamma \gamma$.
%Following our study of unstable vacua, we limit ourselves to low trilinear mass $T_{SO} \lesssim 200-300$ GeV. The constraints on such the pseudo-)scalar octet are very weak and in the following we will consider light pseudo-scalar octet with masses around $900$ GeV.
%Since the pseudo-scalar octet should b Therefore, g

\begin{figure}[!htbp]
\begin{center}
\includegraphics[width=0.6\textheight]{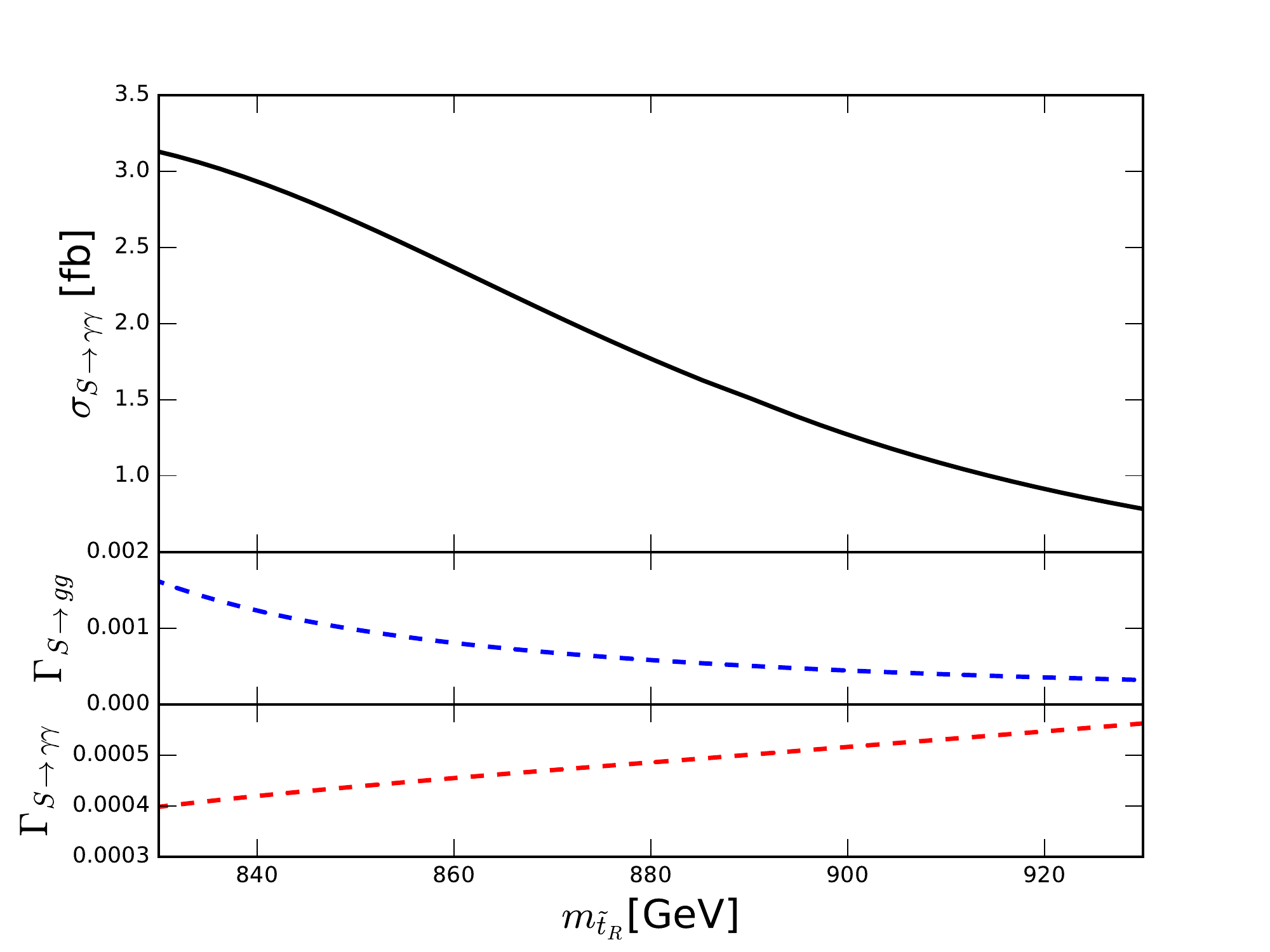}
\caption{$\Stogg$ cross section in $\fb$ as a function of the one-loop mass for the right-handed squarks. The lower two parts show the amplitudes to $\gamma \gamma$ and to $gg$}
\label{fig:S1a_mo}
\end{center}
\end{figure}

Regarding the scalar singlet production, gluon fusion proceeds mainly through loops of $800$ GeV right-handed stop and TeV right-handed first two squarks generations, while left-handed squarks are heavier at $1.75$ TeV. As a consequence, the mass of the right-handed stop is a critical parameter in enhancing $\sigma_{\gamma\gamma}$, we illustrate this dependence in Figure~\ref{fig:S1a_mo} where we plot the $\Stogg$ cross-section as a function of the stop one-loop mass, by varying around the benchmark point of Table~\ref{benchmarkS1}. We see that the cross-section decreases very rapidly with the stop mass.

Regarding the scalar singlet decay to diphoton, it proceeds both through loops of light right-handed sleptons (we consider left-handed sleptons above the TeV) controlled by $\gp m_{1D}$ and loops of fake leptons, $\hE, \hEp,R_u$ and $R_d$ which are controlled by a unified Yukawa $\lambda_{SR} = \lambda_{SE} = 0.7$. Furthermore, the fake sleptons also contribute with couplings controlled by $\gp m_{1D}$. In order to maximise the overall contribution, one has to take care that no cancellations occur between the various contributions (particularly for the D-term-induced couplings, which are proportional to the hypercharge of the scalar participating in the loop). Refering to Table~\ref{table_fields} we see that one possible choice is light $\hE$, $R_d$ and right-handed sleptons and heavier $\hEp$, $R_u$ and left-handed sleptons.

In order to have sizable contributions from the (fake) sleptons, we need a reasonably large singlet Dirac mass $m_{1D} \sim 1250 $ GeV, this has the added benefit that it also enhances the squark contributions to the scalar singlet production rate. On the other hand, it increases the tuning of $\lambda_S$ necessary to have a small mixing and additionally implies that we have either a small $\tan \beta$ or a somewhat large $\mu$ term as can be seen from Eq.~\eqref{EQ:HSRestriction}. 

\begin{figure}[!htbp]
\begin{center}
\includegraphics[width=0.65\textwidth]{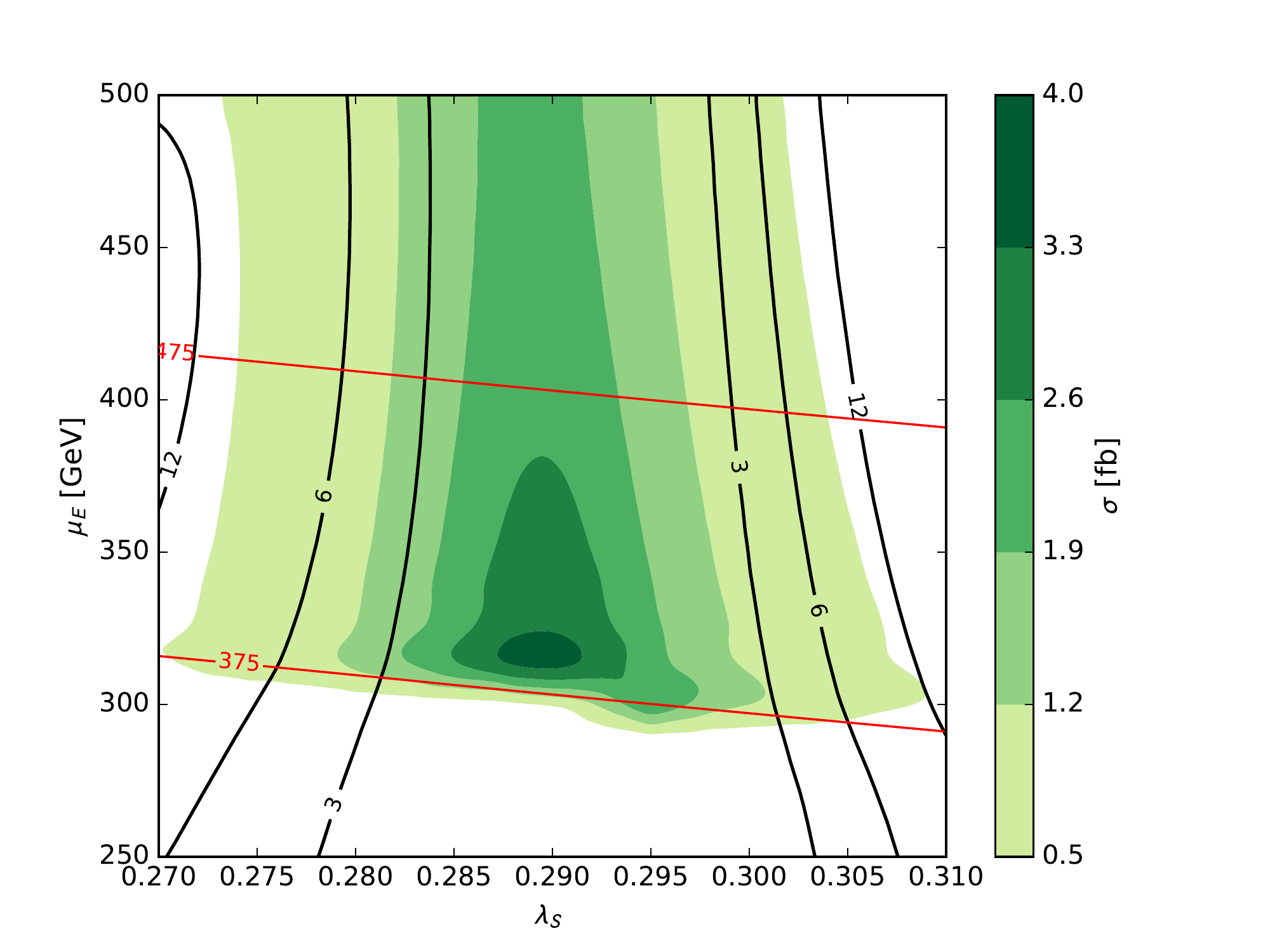}
\caption{$\Stogg$ cross section in $\fb$ as a function of the $\mu_E$ and $\lambda_{S}$. The plot is based on the benchmark point of Table~\ref{benchmarkS1}. The black contour shows the most constraining ratio from~\eqref{RatiosLHC} while the red contours shows the pole mass for the fake leptons.}
\label{fig:S1a_muEls}
\end{center}
\end{figure}

Overall, Figure~\ref{fig:S1a_muEls} presents the cross-section obtained in the $\lambda_S$/$\mu_E$ plane by varying around the benchmark point of Table~\ref{benchmarkS1}. Roughly speaking, this figure combines on the abscissa the constraints from mixing with on the ordinate the requirement that the particles participating in the loop have masses close to half that of the resonance.\footnote{Notice that the fake lepton mass obtains a sizeable contribution from the vev of $S$ throught the $\lambda_{SE}$ term.} 

We see from Figure~\ref{fig:S1a_muEls} that the main requirement in our model is that we must consider values of $\lambda_S$ tuned at the level of a few percent. We can see that the constraint from the ratio $\G_{ZZ} / \G_{\g \g}$ is significantly weaker than the requirement on the cross-section.

\subsubsection*{Octet-induced gluon fusion}

\begin{figure}[!htbp]
\begin{center}
\includegraphics[width=0.65\textwidth]{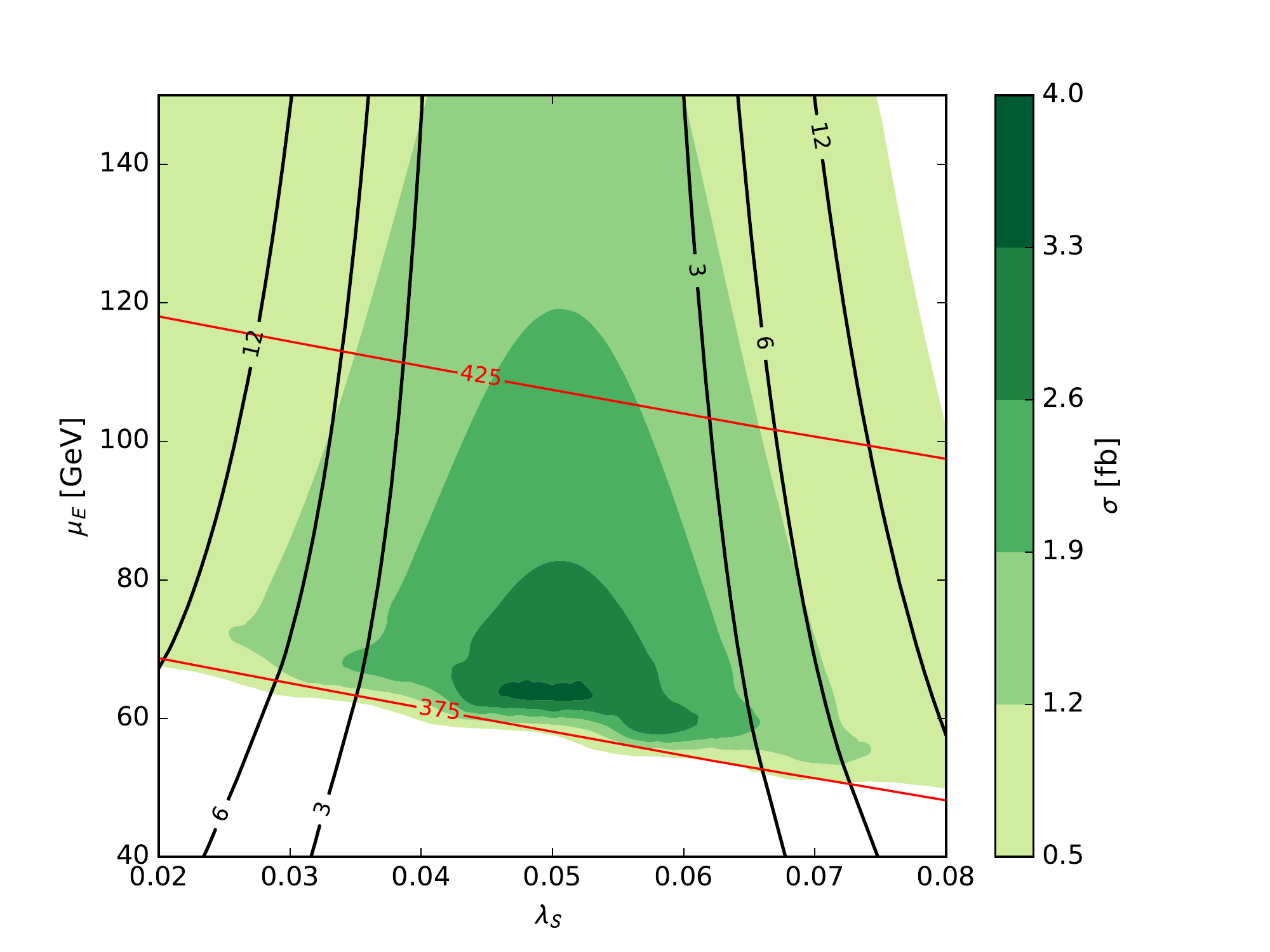}
\caption{$\Stogg$ cross section in $\fb$ as a function of the $\mu_E$ and $\lambda_{S}$ for scenario \Rb. The plot is based on the benchmark point of Table~\ref{benchmarkS1}. The black contour shows the most constraining ratio from~\eqref{RatiosLHC} while the red contours shows the pole mass for the fake leptons.}
\label{fig:S1b_muEls}
\end{center}
\end{figure}

Let us now consider the case  \Rb where we take a light scalar octet. Following the discussion of the previous section, its mass is not constrained as long as we take a large Dirac gluino mass. We will therefore focus on $m_{3D} = 3 $ TeV. Since $m_{3D}$ contributes at tree-level in the mass of the scalar octet, we require large negative $B_O \sim -4 m_{3D}$ in order to have it close to the resonant mass of $375$ GeV. While this is a new source of tuning, the fact that the scalar octet provides a sufficient coupling between the gluons and singlet means that we no longer need a sizeable Dirac mass $m_{1D}$ as in \Ra. As a consequence, the tuning on $\lambda_S$ is milder in this scenario, as can be seen from Figure~\ref{fig:S1b_muEls}. We have presented in Table~\ref{benchmarkS1} a benchmark point for this scenario.

As the singlet Dirac mass is small, the sleptons do not contribute to the singlet decay to diphotons, in stark contrast with scenario \Ra. One relies on loops of fake (s)leptons to increase $\sigma_{\Stogg}$. The crucial parameter in this model is therefore the fake leptons mass, as we illustrate in Figure~\ref{fig:S1b_muEls}.

%% file: Pheno_Rviolating.tex
If we do not constrain ourselves to an R-symmetric scenario, further regions in parameter space open up which can lead to an enhanced di-photon signal. As already stated in Sec.~\ref{sec:prelude}, we discuss mainly two different R-violating scenarios. In the first scenario, we consider the trilinear coupling $\kappa$ still to be small, but allow for the trilinear couplings $T_{SE}$ and $T_{SR}$ as well as for the Yukawa-coupling $\lambda_{SO}$. With the presence of the latter ($\lambda_{SO} = 0.65$), it allows us  to further increase the trilinear coupling $T_{SO}$ (e.g. from $200$ GeV in Scenario \textbf{\Ra} to $1.5$ TeV in Scenario  \textbf{\Rva}) without leading to colour-breaking minima (cf. Sec.~\ref{sec:vacua}). 
\begin{figure}[!htbp]
\begin{center}
\includegraphics[height=0.5\textheight]{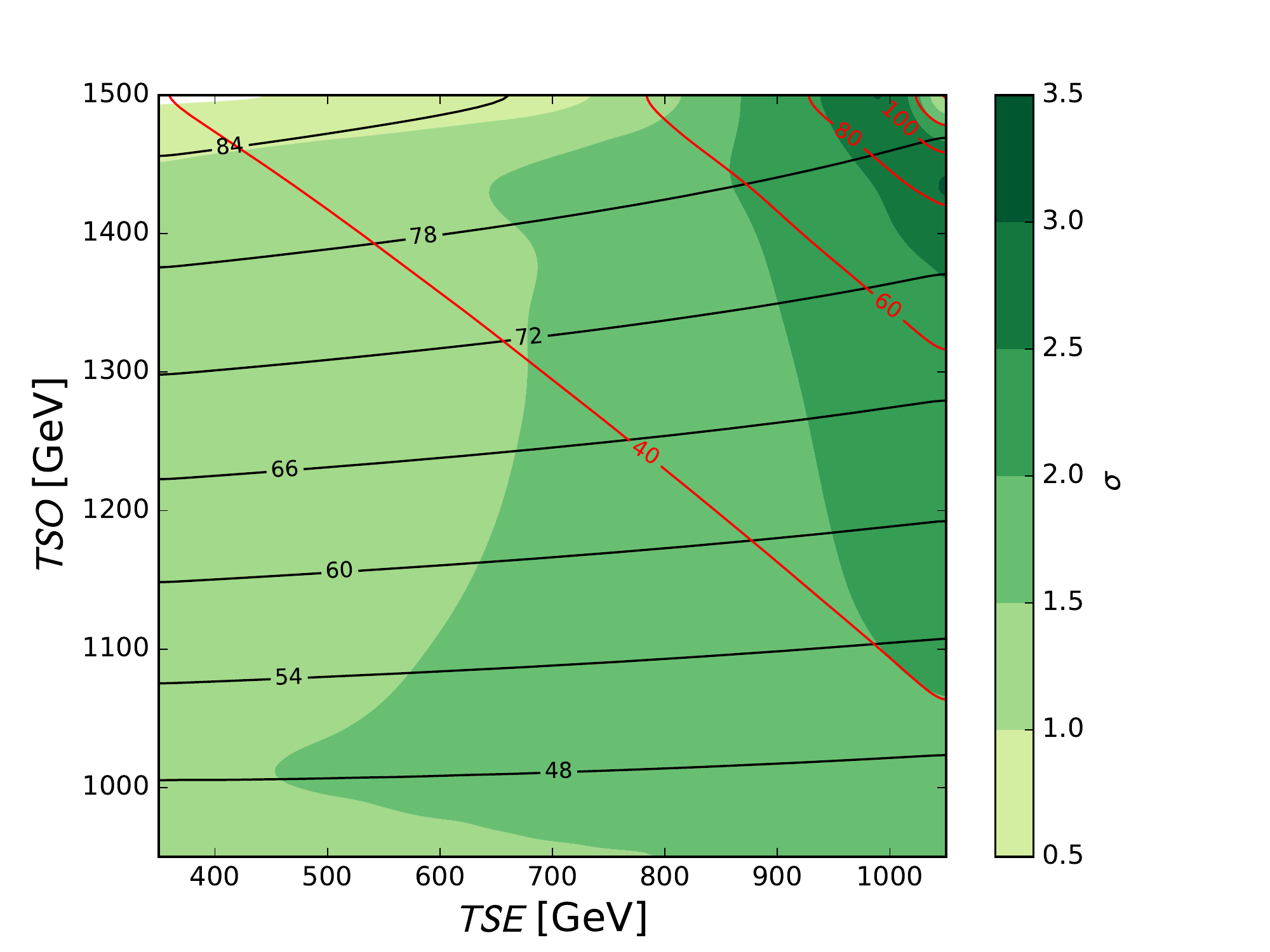}
\caption{Variation of $T_{SE} = T_{SR}$ and $T_{SO}$ while taking the remaining parameters fixed as in benchmark scenario \textbf{\Rva}. $\Gamma_{S \to \gamma \gamma}$ and $\Gamma_{S \to gg}$ are depicted in units of  $\times 10^{-5}$GeV in red and black solid lines, respectively.}
\label{fig:S3_TSO}
\end{center}
\end{figure}
As can be seen in Fig.~\ref{fig:S3_TSO}, the increase of $T_{SO}$ enhances the partial decay width of $\Gamma_{S \rightarrow gg}$, as it increases the coupling of the pseudo-scalar octets ($m_{\sigma_0} = 886.30\, \mathrm{GeV}$) in the loop to the singlet. However, this effect is reduced by an increase of the octet mass via loop effects. At the same time, the increase of $T_{SO}$ leads as well to an decrease of the mass of the fake sleptons, which increases the partial decay width of $\Gamma_{S \rightarrow \gamma \gamma}$. The production via squarks ($m_{\tilde{q}} > 1.7 \mathrm{TeV}$) is suppressed in this scenario. A further enhancement of $\Gamma_{S \rightarrow \gamma \gamma}$ is achieved by increasing $T_{SE} = T_{SR}$.  The fake sleptons and sfermions are below $400 \mathrm {GeV}$ thus leading as well to an enhanced partial decay $S \rightarrow \gamma \gamma$. Similar to the R conserving Scenario,  we account for a mass hierarchy between $\hE$, $R_d$ and $\tilde{e}^c$ and $\hEp$, $R_u$ and $L$ to prevent cancellations from D-term induced couplings, which could still be further increased to allow for an even larger production cross section. Table~\ref{benchmarkRviolating} indicates further input parameters and shows that this benchmark point is in full agreement with current experimental exclusion limits and features a production cross section of $\sigma(S \rightarrow \gamma \gamma) =  3.1$ fb. Choosing a large $m_{3D} = 1600$ GeV, the gluino mass lies above $1800$ GeV, and $m_{T} = 1250$ GeV guarantees that the $\rho$-parameter is below the exclusion limits. We have further checked that such a scenario is not yet excluded by resonant production of $WW$, $ZZ$, $hh$ or $gg$. 

\begin{figure}[!htbp]
\begin{center}
\includegraphics[height=0.5\textheight]{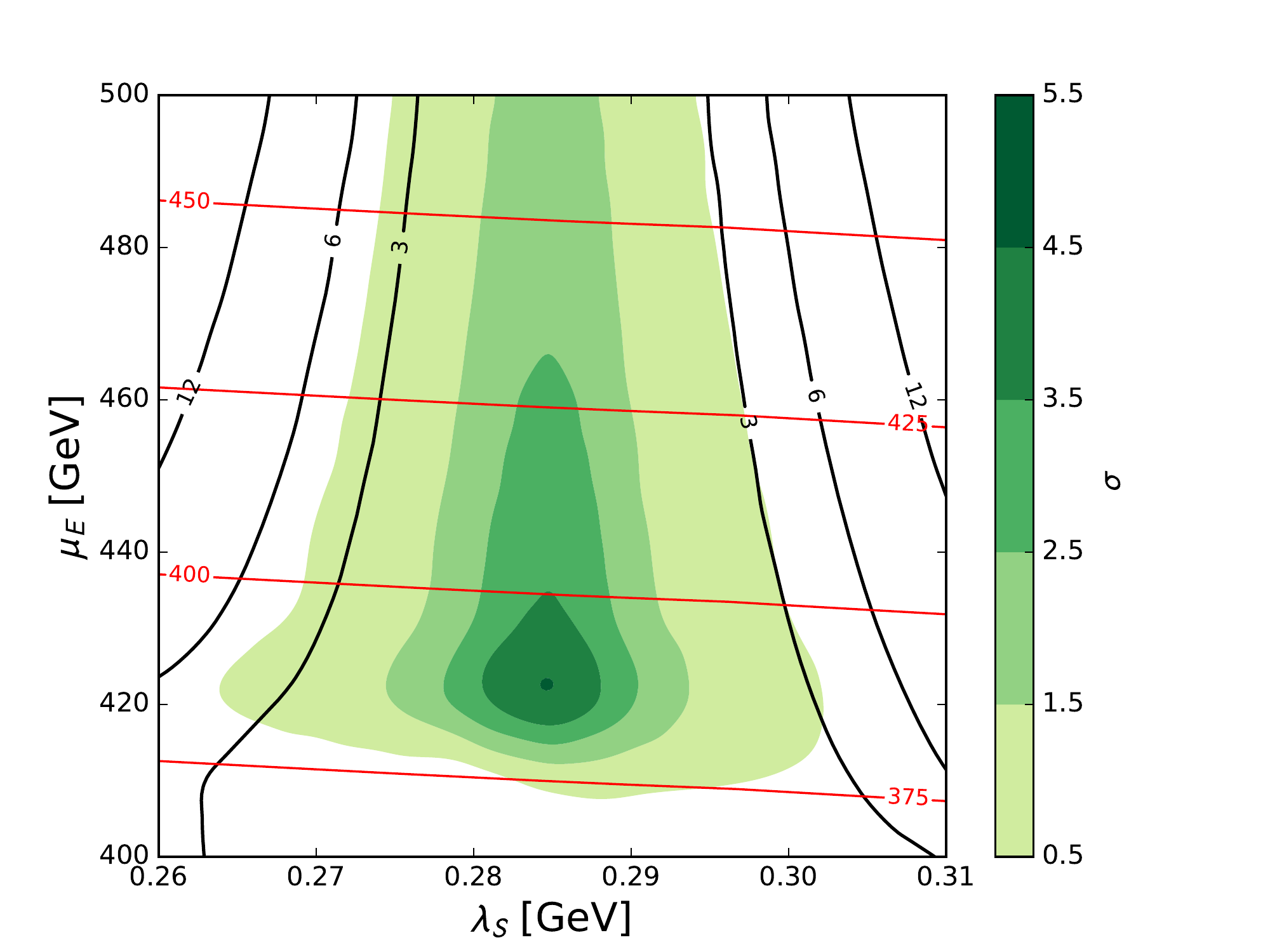}
\caption{$\lambda_S - \mu_E$ plane for benchmark scenario \textbf{\Rva}.}
\label{fig:S3_lamS}
\end{center}
\end{figure}
Fig. \ref{fig:S3_lamS} shows the $\lambda_S - \mu_E$ plane around the benchmark scenario, where a similar behaviour as for the R-conserving scenarios can be observed. Although one would naively expect that in the R-violating scenario the necessary tuning would be less pronounced, the scenario is further constrained when requiring perturbativity of all Yukawa couplings up to the GUT scale. As discussed in Sec. \ref{sec:Landau}, including $\lambda_{SO}$ as free parameter, it further constrains the maximal value of the other Yukawa-couplings. In the studied scenario, a Yukawa-coupling of $\lambda_{SO} \approx 0.65$ implies an upper bound on $\lambda_{SE} < 0.65$ as well. However, generally, larger values of $\lambda_{SO}$ (and respectively of $\lambda_{SE}$) are possible.

\begin{figure}[!htbp]
\begin{center}
\includegraphics[height=0.5\textheight]{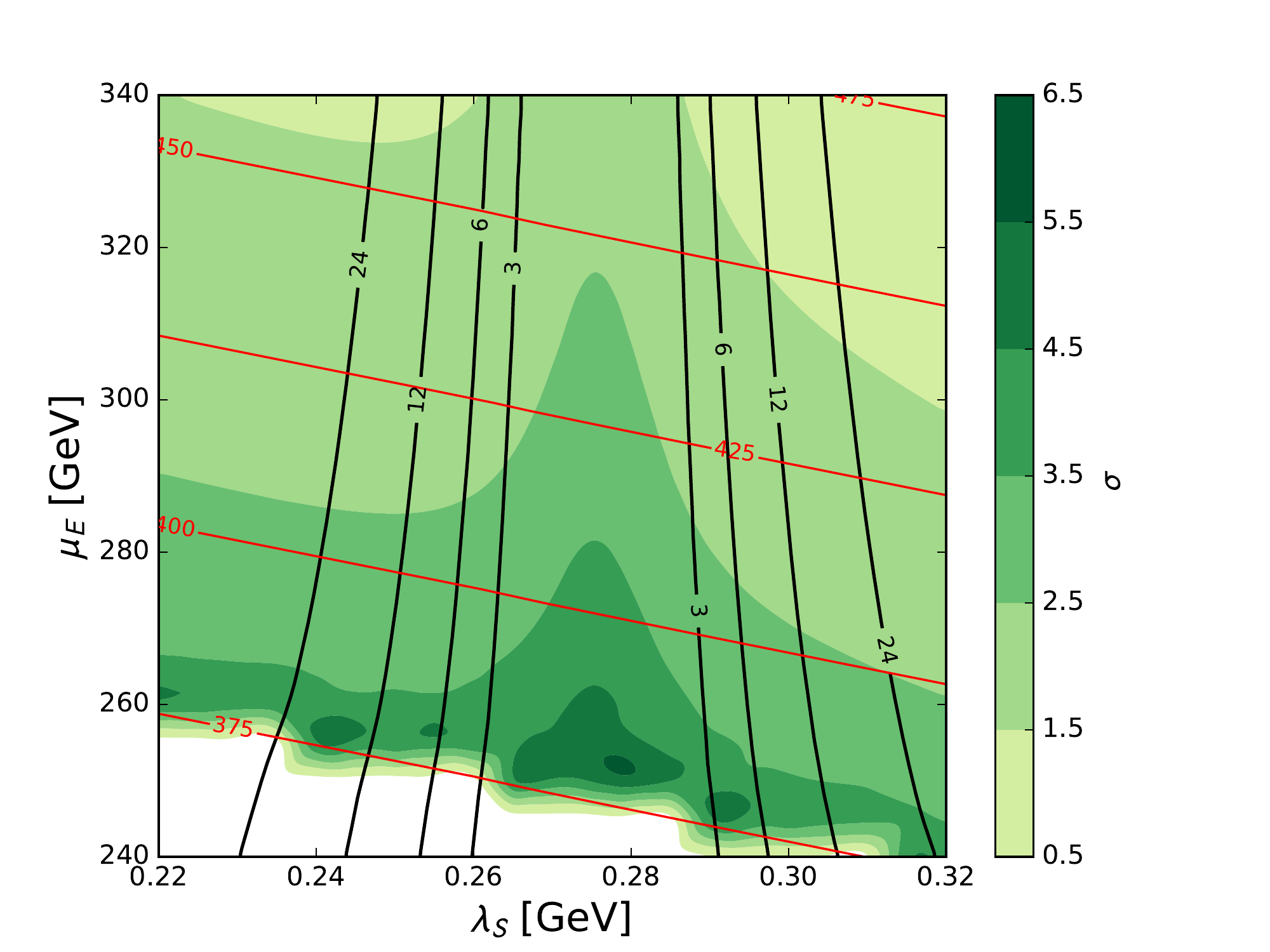}
\caption{$\lambda_S - \mu_E$ plane for benchmark scenario \textbf{\Rvb}.}
\label{fig:S4}
\end{center}
\end{figure}

In the second R violating scenario, we want to discuss the possibility of having a degenerate spectrum of the scalar and pseudo-scalar singlet. Only when the soft SUSY-breaking mass term $M_3$ is included, production via gluon fusion is possible and leads to a sizeable mass splitting between the Majorana gluinos. Having two particles leading to a di-photon signal, a broad parameter space opens up. This can for example be seen in Fig. \ref{fig:S4} showing the corresponding $\lambda_S - \mu_E$ plane around the corresponding benchmark scenario \textbf{\Rvb}. With the pseudo-scalar singlet  not being constrained by mixing with the Higgs like for the scalar singlet, a large variation in $\lambda_S$ is possible as with respect to the R conserving scenarios or \textbf{\Rva}. Due to this degenerate scenario with the pseudo-scalar being less constrained, the requirement of having lower pseudo-scalar octet masses for boosting the production via gluons is loosened ($m_{\sigma_0} = 886.3 \mathrm{GeV}$), as well as for the masses for the fake sleptons and selectrons. 

\begin{table}[!ht]
\begin{center}
\begin{tabular}{c|c|c}
\hline
      Parameter          &  \textbf{\Rva}              & \textbf{\Rvb}\\
\hline
$\mathrm{tan} \beta$   & $2$    &$4$\\
$\mu$   & $660$ GeV  & $450$ GeV\\
$B_{\mu}$ & 2500  GeV  & 2500  GeV\\
$m_{S}$   & $490$ GeV  & $310$ GeV\\
$m_{T}$ & 1250 GeV& 1200 GeV\\
$m_{O}$ &   $530$ GeV &$890$ GeV\\
$M_{3}$ & 0& 1400 GeV\\
$m_{1D}$   & $1250$ GeV  & $490$ GeV\\
$m_{2D}$ & 1000 GeV&  1000 GeV\\
$m_{3D}$ & 1600 GeV&  2300 GeV\\
$\lambda_{S}$   & $0.29$   &$0.27$\\
$\lambda_{T}$   & $0.65$    & $0.70$\\
$\lambda_{SO}$   & $0.65$  &$0.65$\\
$\lambda_{SR}=\lambda_{SE}$   & $0.65$   & $0.65$\\
$B_{S}$   & $-2.4^2$ TeV$^2$ & $-0.7^2$ TeV$^2$\\
$T_{SE} = T_{SR}  $   & $-1000$ GeV    &$0$ GeV \\
$T_{SO}$   & $1500$ GeV &$600$ GeV\\
$M_{Q}$ & 2000 GeV&   2000 GeV\\
$M_{u}$ & 1700 GeV&   1500 GeV\\
$M_{d}$ & 2000 GeV&   2000 GeV\\
$M_{L}$ & 1500 GeV&   1500 GeV\\
$M_{e}$ & 820 GeV&   700 GeV\\
$ \mu_E =  \mu_{R_{u,d}} $    & $413$ GeV & $250$ GeV\\
${m_{\hat{E}}^{11}}^2 = {m_{\hat{E}}^{22}}^2 $   & $400^2$  GeV$^2$ & $400^2$  GeV$^2$\\
${m_{\hat{E}^{\prime}}^{11}}^2 = {m_{\hat{E}^{\prime}}^{22}}^2$   &  $600^2$  GeV$^2$  & $400^2$  GeV$^2$\\
$m_{\hat{R}_{u}}^2$  &  $600^2$  GeV$^2$ & $400^2$  GeV$^2$\\
$m_{\hat{R}_{d}}^2$   & $400^2$  GeV$^2$ & $400^2$  GeV$^2$\\
$B_{\hat{E}}^{11}=B_{\hat{E}}^{22}=B_{\hat{R}}$ & 88500  & 22200\\
\hline
\hline
 $m_h $   &  $124.8$ GeV & $125.9$ GeV\\
  $m_{S} $   & $755.7$  GeV & $756.5$  GeV\\
  $m_{\tilde{S}} $  & $1125.1$  GeV & $751.0$  GeV\\
    \hline
  $m_{\sigma_0} $   & $886.3$  GeV & $886.3$  GeV\\
  $m_{\tilde{e}} $   & $382.2$  GeV & $386.7$  GeV\\
  $m_{e} $  & $378.6$  GeV & $377.2$  GeV\\
    \hline
  $m_{\tilde{u}} $   & $1776.5$  GeV & $1597.2$  GeV\\
  $m_{\tilde{g}} $   & $1825.8$  GeV & $1916.0$  GeV\\
\hline
\hline  
 $ZZ $   &  0.1 & 0.0\\
  $hh$   & 0.5 & 1.2\\
  $WW $   & 0.3 & 0.0\\
  $gg $  & 0.7 & 4.4\\
\hline
\hline
     $\Delta \rho $  &  $9.9 \times 10^{-5}$ & $2.4 \times 10^{-4}$\\
  \small{ $ \sigma ( \Stogg ) $ }    &  \textbf{3.1 fb} & \textbf{4.4 fb}\\
\hline
\end{tabular}
\caption{Overview of the input parameters, physical masses, constraints and production cross section for both R-violating scenarios \textbf{\Rva} and \textbf{\Rva}.}
\label{benchmarkRviolating} 
\end{center}
\end{table}

%% file: Conclusions.tex
The MDGSSM is promising as a model that reproduces the di-photon excess observed at both LHC experiments, ATLAS and CMS. It automatically contains a singlet with both scalar ($S_R$) and pseudoscalar ($S_I$) components that can both be at the origin of the resonance. It is quite easy to fix the mass of one or both of them at $750$ GeV. In the latter case, a small splitting will simulate the larger width of the resonance for which there is a mild preference in the present ATLAS data. Also the model contains new states beyond those of the MSSM, triplets, octets and fake leptons, that can be used in the loops to generate both the production of the singlet and its decay to photons. We have shown that there are diverse experimental constraints that are quite stringent. 

We have found that if the resonance is to be identified with the scalar $S_R$, keeping its mixing with the Standard Model Higgs within the experimentally allowed range represents the most constraining  issue. We have found that a certain amount of cancellation is needed between certain parameters and this can be translated in a tuning of the trilinear $\lambda_S$ at the level of a few percent. Fortunately, this happens to values of  $\lambda_S$ sitting in a quite natural range, near the values expected from an $N=2$ supersymmetric origin of the coupling. 

We found that while remaining within the assumptions of the MDGSSM -- perturbative couplings up to the GUT scale, R-symmetry-breaking only in the superpotential -- the signal can be easily fit by including new dimensionful trilinear couplings of just the adjoints. The latter have not attracted attention in existing literature on Dirac gaugino models, despite the fact that they respect R-symmetry and so are always allowed. While they come out typically small in some scenarios of supersymmetry breaking, this is not always the case and they are expected to be present in the model. We have provided a first comprehensive discussion on this point. We then performed  numerical scans of large parts of the parameter space using the most advanced tools available and in particular the most sophisticated calculation of the Higgs mass (up to two loop order). We found different regions of the parameter space of the MDGSSM (and given example benchmark points) satisfying all existing constraints while providing a good fit to the observed di-photon excess. Moreover, the relevant points have large quantum corrections (in particular to the singlet mass and vacuum expectation value) underlining the importance of using numerical tools.